\documentclass{aa}  

\usepackage[colorlinks=true,breaklinks=true]{hyperref}
\usepackage{graphicx}
\usepackage{txfonts}
\usepackage[dvipsnames,table]{xcolor}
\hypersetup{urlcolor=PineGreen,
	        citecolor=PineGreen,
	        linkcolor=PineGreen}
\usepackage{soul}
\usepackage{caption}
\usepackage{subcaption}
\usepackage[utf8]{inputenc}

\DeclareUnicodeCharacter{03B1}{$\alpha$}

\begin{document}

   \title{Portrait of a Galaxy on FIRE: Is the $\alpha$-bimodality a natural consequence of inside-out disc growth in a hierarchical formation scenario?
   }

   \author{
        María Benito\inst{1, 2, 3}\thanks{e-mail: mariabenitocst@gmail.com} \and
        Annaliina Aavik\inst{1} \and
        Giuseppina Battaglia\inst{2, 3} \and
        Salvador Cardona-Barrero\inst{3, 2} \and
        Ele-Liis Evestus\inst{4} \and
        Emma Fernández-Alvar\inst{3, 2} \and
        Sven Põder\inst{4, 5} \and
        Heleri Ramler\inst{1} \and
        Boris Deshev\inst{1} \and
        Elmo Tempel\inst{1}
        }

   \institute{
    Tartu Observatory, University of Tartu, Observatooriumi 1, Tõravere 61602, Estonia. \and
    Instituto de Astrofísica de Canarias, Calle Vía Láctea s/n E-38206 La Laguna, Santa Cruz de Tenerife, España. \and
    Universidad de La Laguna, Avda. Astrofísico Francisco Sánchez E-38205 La Laguna, Santa Cruz de Tenerife, España. \and
    Tallinn University of Technology, Ehitajate tee 5, Tallinn 19086, Estonia \and
    National Institute of Chemical Physics and Biophysics (NICPB), Rävala 10, Tallinn 10143, Estonia
    }

   \date{Received \today; accepted}
 
  \abstract 
   {
   The chemical dichotomy in the [$\alpha$/Fe]-[Fe/H] plane is a consequence of the complex processes underlying the formation and evolution of disc galaxies such as observed in the stellar Milky Way disc.
   }
   {We determine what can drive an $\alpha$-bimodality of the disc in a zoom-in hydrodynamical simulated galaxy which has had no major mergers and negligible radial migration. 
   }
   {Using a Milky Way-mass galaxy from the FIRE-2 suite of simulations, we analyse gas flows in the disc together with its star formation and merger history, as well as the chemical evolution of the hot corona, to investigate their connection to transitions in the chemo-dynamical structure of the stellar disc and its radial distribution.}
   {The simulated galaxy exhibits high and low-$\alpha$ sequences without having experienced major mergers nor significant radial migration. 
   A high-$\alpha$ thick disc forms during the early chaotic clustering phase. Afterwards, as the star formation rate declines, a dip in the stellar number density appears, coinciding with the dilution of the galactic corona by a minor merger, which subsequently halts the rise of [Fe/H] in the disc. Later, accreted gas onto the disc from minor mergers, mildly enhances the star formation rate and generates the low-$\alpha$ sequence in the outer disc, with radial inward flows of this material feeding the low-$\alpha$ inner disc. 
   Furthermore, we find that even at fixed radii, newly formed stars retain a sizable spread in their chemical abundances, reflecting chemical differences between the in-situ and the infalling gas from which they formed, further indicating that instantaneous gas mixing is invalid.
   }
   {Understanding the chemical evolution of stellar discs requires accounting for their accretion merger history and interaction with the surrounding hot corona, as well as the vertical and radial gas flows that redistribute metals within the disc.
   }

   \titlerunning{Portrait of a Galaxy on FIRE}
   \authorrunning{Benito et al.}
   \maketitle

\section{Introduction}

The Milky Way (MW) stellar disc exhibits a well-defined chemical bimodality, with stars occupying two distinct sequences in the [$\alpha$/Fe]-[Fe/H] plane~\citep{1998A&A...338..161F, 2014A&A...562A..71B, 2015ApJ...808..132H, 2023ApJ...954..124I, 2024A&A...682A...9G}. The stars in each of these sequences show different kinematic and spatial trends that represent the end product of the Galaxy's formation and evolution. Understanding the origin of these features can therefore constrain the physical processes that shape the evolution of galactic discs, including gas flows, radial migration, and feedback mechanisms.

Several scenarios have been proposed to explain the observed bimodality. For instance, infall chemical evolution models argue that distinct episodes of gas accretion and star formation can establish different abundance sequences (e.g., \citealt{1997ApJ...477..765C, 2022A&A...663A.174S}). 
However, recent work has found that infall scenarios of pristine gas face strong constraints, as they must reproduce the $\alpha$-bimodality without excessive dilution of the metallicity of the interstellar medium, since such dilution would prevent these models from matching the observed age–abundance trends \citep{2025arXiv250800988D}.
Another hypothesis emphasises the role of stellar radial migration, in which 
stars originating in regions with different chemical enrichment histories can mix throughout the disc due to changes in the stars' specific angular momentum through non-axisymmetric forces, such as those imparted by the bar or spiral arms
(e.g., \citealt{2002MNRAS.336..785S, 2009MNRAS.396..203S, 2013A&A...558A...9M}). 
A third scenario explains the chemical dichotomy as the outcome of two distinct modes of star formation. 
On the one hand, in the model of \cite{2019MNRAS.484.3476C}, the formation of the high and low-$\alpha$ sequences is not sequential. The former forms in high-density star formation rate (SFR) clumps that dominate the early galaxy, and the latter forms in spatially extended regions of low SFR density.
On the other hand, \cite{2021MNRAS.501.5176K} showed that two star formation regimes arise naturally in isolated MW–like simulated galaxies as a consequence of an initially rapid dissipative gas collapse that drives a high SFR, builds the high-$\alpha$ sequence, and triggers strong stellar feedback that expels gas from the disc. The subsequent re-accretion of this gas on longer time-scales sustains a more quiescent star formation phase, giving rise to an extended low-$\alpha$ sequence. These two regimes are supported with observational reconstructions of the MW’s star formation history (e.g. \citealt{2014ApJ...781L..31S}).

Furthermore, the prevalence of $\alpha$-bimodalities in external disc galaxies is also unknown. For instance, whether a chemical bimodality exists in M31 remains a matter of debate, as currently available abundance measurements are sparse and provide inconclusive or model-dependent evidence for different $\alpha$ sequences~\citep{2023ApJ...956L..14K, 2024IAUS..377..115N}. From a theoretical point of view, the identification of bimodalities in simulations depends on how the bimodality is defined in abundance space and on the adopted analysis methodology, with no currently agreed, consistent criterion applied across different simulation suites. As a result, it remains unclear whether this feature is a universal outcome of MW-mass galaxy formation models (e.g., \citealt{2018MNRAS.474.3629G, 2018MNRAS.477.5072M, 2025MNRAS.537.1571P}). All this raises the question of whether the $\alpha$-bimodality is a ubiquitous feature of disc galaxies or an outcome of the particular formation history of the Milky Way, highlighting the importance of establishing its presence in external systems (e.g., \citealt{2024A&A...691A..61P}).

In this paper, we analyse the role of gas accretion and radial gas flows in creating a chemical dichotomy in a simulated hydrodynamical, zoom-in MW-mass galaxy. We quantify the fluxes of gas across the disc, examine their relation to changes in the star formation and merger history, their origin and interplay with the galactic corona, and investigate how they contribute to the observed chemical and kinematic evolutionary trends of the stellar disc. For this purpose, we select a simulated MW-mass galaxy with a clear $\alpha$-bimodality, which at present has no bar and has had a quiet merger history. The layout of this paper is as follows: 
in Section~\ref{sec:Romeo}, we describe the general properties of the simulation we use in this paper. Section~\ref{sec:Disc} details the spatial, kinematical and chemical properties of its stellar disc. 
Section~\ref{sec:chemical-evol} then discusses the physical mechanisms that drive these properties and, specifically, the role of gas accretion and radial gas flows, along with the chemical evolution of the galactic corona and {\sc Romeo}'s merger history. Lastly, we discuss our results and present the conclusions in Section~\ref{sec:conclusions}.

\section{\textsc{Romeo} in context}
\label{sec:Romeo}
\subsection{Simulation details}

We analyse the simulated \textsc{Romeo} galaxy \citep{2019MNRAS.487.1380G}, part of the ELVIS (Exploring the Local Volume in Simulations) FIRE-2 cosmological zoom-in suite \citep{2018MNRAS.480..800H}, which models galaxy pairs similar to those in the Local Group. The centre of {\sc Romeo} is defined as in~\cite{wetzel_public_2023}. An iterative zoom-in approach is applied to the star particles: starting from their mean centre-of-mass, the enclosing sphere is repeatedly shrunk by 50\% until its radius falls below $\sim$10 pc. The systemic velocity is then taken as the centre-of-mass velocity of star particles within 8 kpc.

\textsc{Romeo} is the more massive of the pair, and its disc formed earlier than those of the other galaxies in the suite. In particular, it started forming at 11 Gyr in lookback time, at odds with other FIRE-2 galaxies, where disc settling took place from about 9 Gyr ago onwards~\citep{mccluskey_disc_2024}, but comparable with the Milky Way disc, which is thought to have settled $10-13 \,\rm Gyr$ ago~\citep{2003A&A...410..527B, 2013A&A...560A.109H, 2022MNRAS.514..689B, 2022Natur.603..599X, gallart_chronology_2024}. As we will see, the stellar disc in {\sc Romeo} exhibits an $\alpha$-bimodality, which is not unusual within FIRE-2 galaxies since 8 (including {\sc Romeo}) of the 11 simulated galaxies analysed by \cite{2025MNRAS.537.1571P} were identified as having a bimodality. We note, however, that the number of galaxies identified as exhibiting $\alpha$-bimodality in \cite{2025MNRAS.537.1571P} might be higher, since the definition of bimodality adopted is conservative.
In addition, {\sc Romeo} only develops a bar during a short period at the very end of its evolution~\citep{2025ApJ...978...37A}, after the onset of the $\alpha-$bimodality. This means that most of its evolution is unaffected by bar-driven radial mixing of stars, allowing us to isolate the role of gas flows in shaping the chemical evolution of its stellar disc.

The simulation was run using the GIZMO code in mesh-less finite-mass (MFM) mode \citep{hopkins_new_2015} and using the FIRE-2 galaxy formation model \citep{hopkins_fire-2_2018}. An overview of the details of this model can be found in \citet{hopkins_fire-2_2018, wetzel_public_2023}. {\sc Romeo} is simulated in a flat $\Lambda$CDM cosmology with parameters $\Omega_m = 0.31$, $\Omega_{\Lambda} = 0.69$, $\Omega_b = 0.048$, $h = 0.68$, $\sigma_8 = 0.82$, $n_s = 0.97$. 
Dark matter particles have a mass of $M_{DM} = 1.9\times 10^4 M_{\odot}$, and the initial mass of star particles and gas cells is $M_{bar} = 3.5 \times 10^3 M_{\odot}$. Star particles lose mass over time through stellar winds, which transfer mass to the gas. Dark matter and stars have gravitational softening lengths $\epsilon_{DM} = 32$ pc and $\epsilon_{star} = 4.4$ pc, respectively. The softening length of the gas is adaptive, and reaches a minimum value of $\epsilon_{gas,min} = 0.7$ pc. The simulation consists of 600 snapshots in the range $z = 99$, $t = 0$ Gyr to $z = 0$, $t\simeq13.73$ Gyr with time spacing $\lesssim 25$ Myr. In our analysis, we used the 39 snapshots that are publicly available and have an average and maximum time spacing of $\sim 0.3$ Gyr and $1.57$ Gyr, respectively. 

\subsection{General properties}

Table~\ref{tab:global} displays several global characteristics of the \textsc{Romeo} simulated galaxy and the Milky Way. While the virial DM masses\footnote{DM mass within a sphere in which the average DM density equals 200 times the critical density of the universe at a given z.} of both galaxies are similar within observational uncertainties, \textsc{Romeo} exhibits greater stellar and gas masses compared to the Milky Way. However, it is important to note that the MW DM halo mass remains uncertain, within a factor of at least two~\citep{2016ARA&A..54..529B, 2020JCAP...05..033K}.

\textsc{Romeo} recently experienced a brief bar episode from roughly 900 to 100 Myr ago~\citep{2025ApJ...978...37A}, but since the $\alpha$-bimodality is already established before the bar’s onset, our analysis focuses on the period preceding its formation, that is in the first 12.39 Gyr of the galaxy up to redshift $z\sim0.1$. Table~\ref{tab:global} includes values for both $z\sim0.1$ and $z = 0$, showing that the global properties of \textsc{Romeo} remain similar between these two epochs. This indicates that extending the time frame to include the full 13.73 Gyr evolution of the galaxy does not alter the overall picture. In addition, \textsc{Romeo} has had a quiet merger history with no major mergers. Specifically, it did not undergo any mergers with galaxies exceeding 1\% of its own stellar mass after $z = 3$, with the ratio calculated as the peak satellite stellar mass over the stellar mass of the host at the time of the merger.

\begin{figure}
    \centering
    \includegraphics[width=0.38\textwidth]{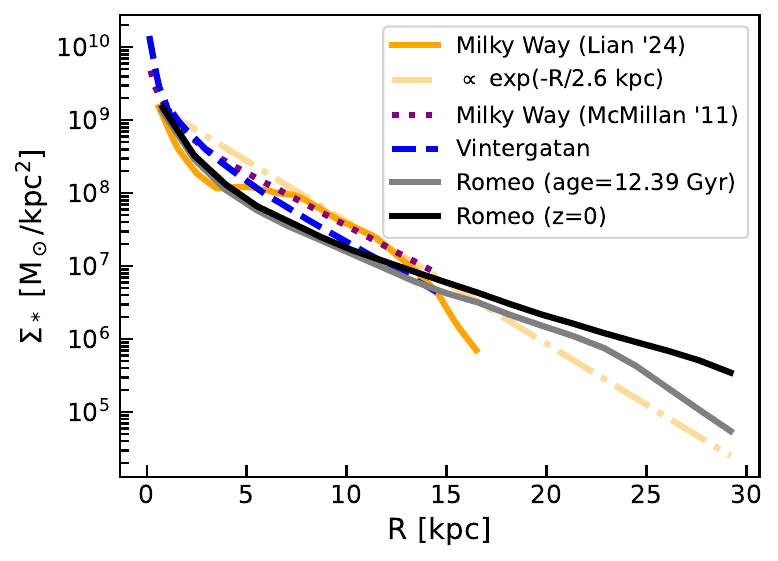}
    \includegraphics[width=0.38\textwidth]{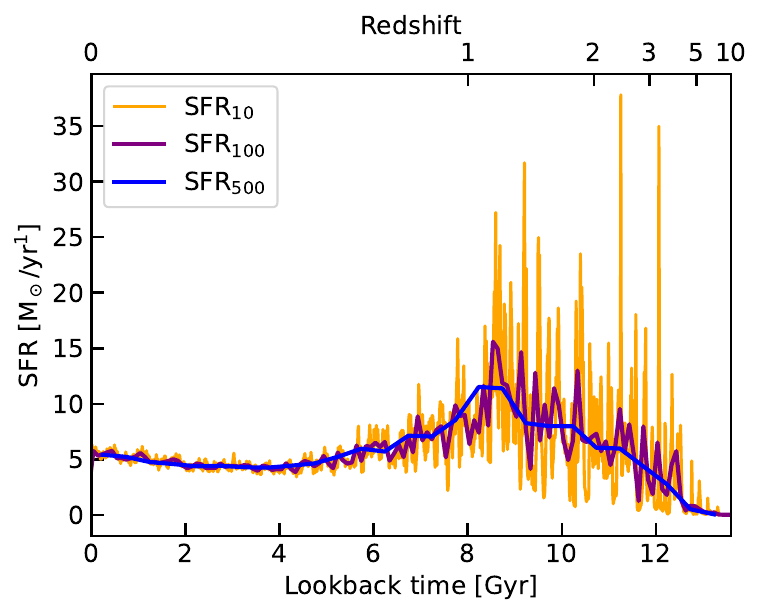}
      \caption{Top: Stellar surface density profiles of the Romeo and Vintergatan~\citep{2021MNRAS.503.5826A} simulated galaxies, together with the Milky Way profile taken from \cite{2024NatAs...8.1302L} and the best-fitting morphology of \cite{2011MNRAS.414.2446M}. The former Milky Way profile has been normalised to enclose a stellar mass of $4\times10^{10}\,\rm M_\odot$. The dot-dashed orange line depicts a falling exponential profile with scale-length of 2.6 kpc.
      Bottom: Star formation rate as a function of time in \textsc{Romeo} calculated with timesteps $\Delta t = 10$ Myr (orange), $\Delta t = 100$ Myr (purple) and $\Delta t = 500$ Myr (blue).
    }\label{fig:general_properties}
\end{figure}

The top panel of Fig.~\ref{fig:general_properties} compares the stellar surface density profile of \textsc{Romeo} with those of the Vintergatan simulated galaxy~\citep{2021MNRAS.503.5826A} and the observed Milky Way. 
\textsc{Romeo} exhibits the most extended disc with an exponential scale-length of 4.5 kpc at $z = 0$ and 4.1 kpc at a lookback time of 1.34 Gyr (right before the bar period). This is larger than the Milky Way's estimated value of 2.6 kpc~\citep{2016ARA&A..54..529B}, or the value of $3.26\pm0.25$ kpc, most recently estimated using Gaia DR3~\citep{2025A&A...701A.270K}. {\sc Romeo} is also less concentrated as evident from \textsc{Romeo}'s lower surface density up to $\sim 15$ kpc. Similar to the Milky Way (see solid orange line in top panel of Fig.~\ref{fig:general_properties} and ~\citealt{2024NatAs...8.1302L}), \textsc{Romeo} has a broken surface density profile at 1.34 Gyr in lookback time, although the break radius occurs at larger galactocentric distances.

\renewcommand{\arraystretch}{1.3}
\begin{table}[t!] 
\caption{Global properties of the \textsc{Romeo} simulated galaxy at $z\sim0.1$, just before the start of its bar episode, and at $z=0$, compared with those of the Milky Way.}
\centering
\small\addtolength{\tabcolsep}{1pt}
\resizebox{\columnwidth}{!}{%
\begin{tabular}{  l  c  c  c  c  c }
    \hline
     & $z$ & $M_{200}^{\rm DM}$ & $R_{200,c}$ & $M_{\rm gas}$ & $M_*$ \\
     & & $[10^{11}\, \rm M_{\odot}]$ & $[\rm kpc]$ & $[10^{10}\,\rm  M_{\odot}]$ & $[10^{10}\,\rm  M_{\odot}]$ \\
    \hline
    \textsc{Romeo} & 0.0998 & $9.2$ & $200$ & $8.4$ & $7.4$ \\
    \hline
    \textsc{Romeo} & 0 & $9.4$ & $206$ & $8.3$ & $8.0$ \\
    \hline
    Milky Way & 0 & $8.3^{+1.0}_{-0.9}$ & $193\pm 7$ & $\sim 5$ & $5\pm1$ \\
    \hline
\end{tabular}}
\tablefoot{The virial radius, $R_{200, c}$, is defined as the radius of the sphere in which the average DM density equals 200 times the critical density of the universe at given $z$. The masses $M_{200}^{\rm DM}$, $M_{\rm gas}$, and $M_*$ are the DM, gas, and stellar masses enclosed within that sphere. The DM, gas, and stellar mass point estimates, with corresponding $1\sigma$ uncertainties, of the Milky Way are taken from~\cite{2020JCAP...05..033K}, \cite{2015ApJ...800...14M} and \cite{2016ARA&A..54..529B}, respectively.
}\label{tab:global}
\end{table}

\renewcommand{\arraystretch}{1.3}
\begin{table*}[t!] 
\caption{Mean rotational velocity $\bar{v}_\phi$ and velocity dispersions in the cylindrical $R$ and $z$ coordinates for the high-$\alpha$/bridge/low-$\alpha$ sequences.}
\centering
\small\addtolength{\tabcolsep}{1pt}
\resizebox{\textwidth}{!}{%
\begin{tabular}{  l  c  c  c  c  c c c}
    \hline
     & $z$ & $\bar{v}_\phi$ & $\sigma_R$ & $\sigma_z$ & $r_d$ & $h_z$ & $M_*$\\
     & & $[\rm km/s]$ & $[\rm km/s]$ & $[\rm km/s]$ & $[\rm kpc]$ & $[\rm kpc]$ & $[10^{10}\,\rm  M_{\odot}]$ \\
    \hline
    \textsc{Romeo} & 0.0998 & 150/219/233 & 124/70/42 & 97/44/22 & 2.4/4.6/7.3 & 1.7/0.6/0.3 & 0.9/1.9/0.7 \\
    \hline
    \textsc{Romeo} & 0 & 160/218/230 & 116/70/47 & 104/40/21 & 2.2/4.6/7.5 & 1.2/0.6/0.3 & 0.9/1.9/1.4 \\
    \hline
    Milky Way & 0 & 194/214/228  & 63/48/37 & 42/29/20 & $2.0\pm0.2/2.6\pm0.5$ & $0.9\pm0.2/0.3\pm0.1$ & $0.6\pm0.3/3.5\pm1$ \\
    \hline
\end{tabular}}
\tablefoot{
For the Milky Way, these kinematic quantities were derived from a sample of giants in Gaia DR3 crossmatched with APOGEE DR17 (see text for details). The corresponding same values for \textsc{Romeo} are calculated using the 3 stellar discs components identified via GMM (see section ~\ref{subsec:GMM} for details). We also include the scale-length, scale-radius and stellar masses for these components, which can be compared with the values inferred for the thick($\approx$high-$\alpha$)/thin($\approx$low-$\alpha$) discs of the Milky Way~\citep{2016ARA&A..54..529B}.
}\label{tab:disc}
\end{table*}

The bottom panel of Fig.~\ref{fig:general_properties} depicts the global star formation history of {\sc Romeo}. We determine the SFR by summing the masses of stars born within a cylinder of radius $R < 40{\, \rm kpc}$ and height $|z| < 10{\, \rm kpc}$ at a chosen timestep $\Delta t$, with the mass of each star measured at the moment of its birth.
To evaluate the burstiness of the SFR, we use $\Delta t = 10$ Myr, as this corresponds to the typical snapshot time spacing \citep{wetzel_public_2023}, providing the best quantification of the instantaneous SFR without being affected by stochasticity. For a direct comparison with the Vintergatan simulated galaxy~\citep{2021MNRAS.503.5826A}, we adopt $\Delta t = 100$ Myr. Finally, to smooth the SFR and capture the overall trend, we use $\Delta t = 500$ Myr. The star formation history (SFH) of Romeo exhibits an early bursty phase characterised by an overall rise in the SFR, followed by quiescent SF at later times when the SFR declines, with a mild increase closed to the present epoch. These phases in the SFH are directly connected to the different evolutionary stages of the disc and the development of the $\alpha$-bimodality, as we will discuss below, and broadly resemble the behaviour seen in the Vintergatan simulation.

\section{Disc properties}
\label{sec:Disc}
\subsection{Gaussian Mixture modelling of components}
\label{subsec:GMM}
We note that the term disc is used at two levels in this work. At a global level, it refers to the entire stellar disc of the simulated galaxy, {\sc Romeo}. At a more fine-grained level, the global stellar disc is subdivided into distinct components purely based on differences in the age and kinematical properties of its star particles. To identify these disc components, we first performed a Gaussian Mixture Model (GMM) analysis, which allows to classify stellar particles in a probabilistic manner, avoiding hard cuts and being agnostic about the number of distinct components (see also~\citealt{2021ApJ...921..106N}). Thus enabling a direct comparison of the best-fit model with traditional kinematic and stellar-abundance divisions.

We applied GMM using 3D cylindrical velocities and the ages of star particles within $R < 40{\, \rm kpc}$ and $|z| < 10{\, \rm kpc}$ at $z=0$. That is, the probability density of these four properties for each individual star particle in the volume of interest, $\boldsymbol{d}_i=(v_R, v_\phi, v_z, A)_i\in \mathbb{R}^4$, is given by a sum of $N$ Gaussians:
\begin{equation}
   p(\boldsymbol{d}_i\,|\, \{\boldsymbol{e^w}_n, \boldsymbol{\mu}_n, \boldsymbol{\Sigma}_n\}_{n=1}^N) =\sum_{n=1}^N \boldsymbol{e^w}_n \,\mathcal{N}(\boldsymbol{d}_i \, | \, \boldsymbol{\mu}_n, \boldsymbol{\Sigma}_n),
\label{eq:GMM_like}
\end{equation}
with weight $\boldsymbol{e^w}_n\in\mathbb{R}^4$, mean $\boldsymbol{\mu}_n\in\mathbb{R}^4$ and covariance matrix $\boldsymbol{\Sigma}_n\in\mathbb{R}^{4\times 4}$. Including stellar age alongside cylindrical velocities allows particles with similar present-day orbital properties to be distinguished by their formation time, thereby enabling the separation of populations formed by different channels.

We allow $N$ to vary from 1 to 16 and used the elbow rule and the deceleration in the decline of Bayesian Information Criterion values to identify $N=4$ as the optimal number of components (see App.~\ref{app:GMM_discs}). For each GMM identified component, we assigned star particles based on their probability of belonging to a specific Gaussian cluster. In particular, star particles in cluster $k$ satisfy:
\begin{equation}
    \frac{\boldsymbol{\hat{e^w}}_k \,\mathcal{N}(\boldsymbol{d}_i \, | \, \boldsymbol{\hat{\mu}}_k, \boldsymbol{\hat{\Sigma}}_k)}{\sum_{n=1}^4 \boldsymbol{\hat{e^w}}_n \,\mathcal{N}(\boldsymbol{d}_i \, | \, \boldsymbol{\hat{\mu}}_n, \boldsymbol{\hat{\Sigma}}_n)} \geq 0.68,
\end{equation}
where $\boldsymbol{\hat{e^w}}_n$, $\boldsymbol{\hat{\mu}}_n$ and $\boldsymbol{\hat{\Sigma}}_n$ are the estimated weight, mean and covariance matrix of each of the four Gaussian components.
These values are inferred by maximising the likelihood given by equation~\eqref{eq:GMM_like}. As the initial values for the parameters are assigned randomly, it is necessary to perform multiple initialisations until a consistent result is reached. We found that seven initialisations were required to reach a stable result for our dataset.

\begin{figure}
    \centering
    \includegraphics[width=0.4\textwidth]{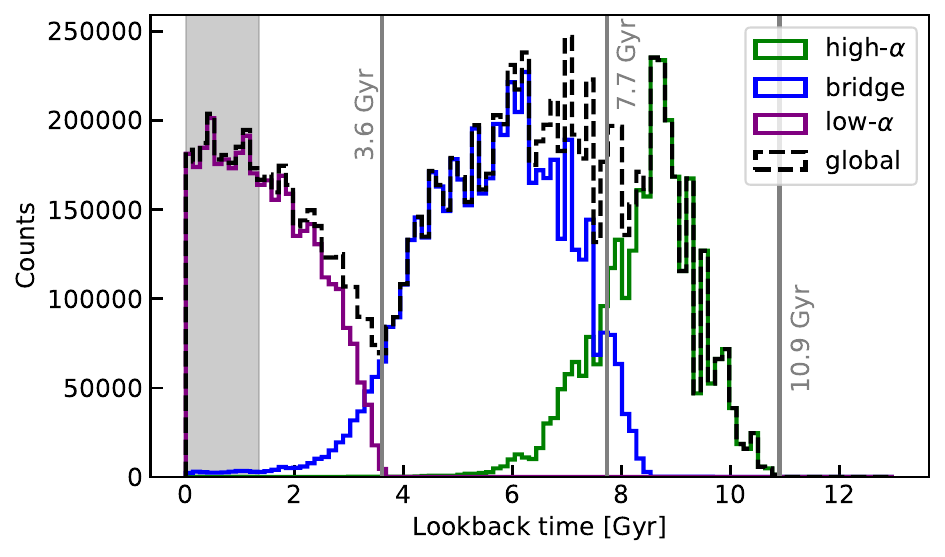}
      \caption{
      One-dimensional age distribution of all disc star particles along with the disc components identified by GMM. The vertical grey lines mark key evolutionary stages of the disc identified by GMM (see Sect.~\ref{subsec:discs}): the onset of the thick or high-$\alpha$ disc (10.9 Gyr ago), the time when the bridge or intermediate-$\alpha$ disc begins to dominate over thick disc (7.7 Gyr ago), and the onset of the $\alpha-$bimodality (3.6 Gyr ago).
      The vertical grey bands mark the timing of the bar episode.
      }
    \label{fig:global_disc_1Dages}
\end{figure} 

\begin{figure}
    \centering
    \includegraphics[width=\columnwidth]{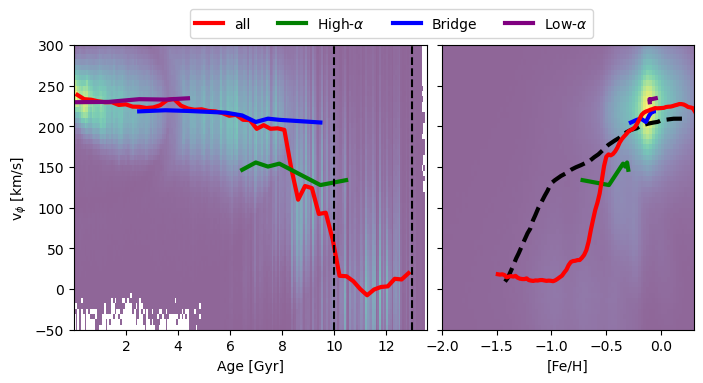}
      \caption{Two-dimensional distribution of the azimuthal or rotational velocity of all stellar particles in {\sc Romeo} $R < 40{\, \rm kpc}$ and $|z| < 10{\, \rm kpc}$ as a function of age (left panel) and metallicity [Fe/H] (right panel), with the median trend shown in red. The green, blue and purple lines depict the median trend of the stellar population in the high-$\alpha$, bridge and low-$\alpha$ discs, respectively. In the left panel, the dashed vertical lines indicate the inferred period at which the Milky Way's stellar disc settles (see Sect.~\ref{subsec:discs}), while the dashed black line in the right panel shows the inferred trend for the Milky Way as in~\cite{belokurov_dawn_2022}.
    }\label{fig:GMM_discs_kinematics}
\end{figure}

After visual inspection of the kinematics of each identified component, we concluded that GMM identifies a halo and three disc-like structures, which, as we shall see in Section~\ref{subsec:discs}, can be associated with high, intermediate and low-$\alpha$ components of the stellar disc.  
The age distribution of each of the three discs is shown in Fig.~\ref{fig:global_disc_1Dages}, and average kinematical properties, along with MW's disc properties, are reported in Table~\ref{tab:disc}. In addition, Fig.~\ref{fig:GMM_discs_kinematics} depicts the distribution of rotational velocities as a function of age and metallicity. As expected, the average rotational velocity decreases with increasing age of the component. Surprisingly, the coherent rotation of the high-$\alpha$ disc increases in the last 1.34 Gyr of evolution of the galaxy, the period within which a weak bar develops in {\sc Romeo}. This together with the change in the vertical distribution of this profile during the bar development in Romeo (see bottom panel of Fig.~\ref{fig:disc_time_evol}) may be driven by the bar. 

\subsection{Disc components}
\label{subsec:discs}
\begin{figure}
    \centering
    \includegraphics[width=0.7\columnwidth]{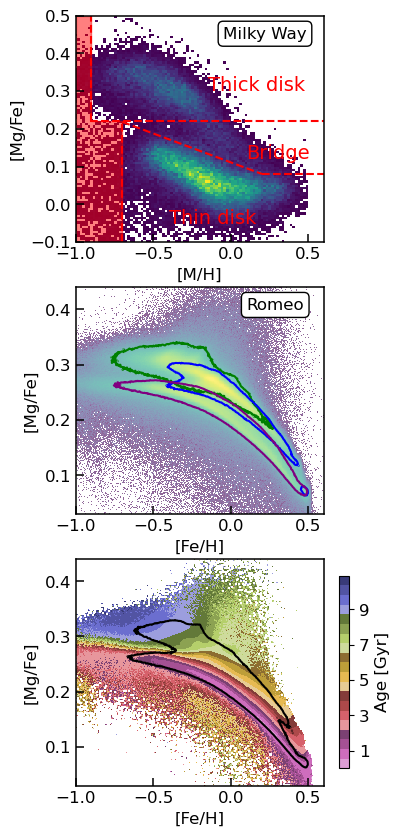}
      \caption{Top panel: two-dimensional number density of a sample of giants in Gaia DR3 crossmatched with APOGEE DR17 that spanned a Galactocentric range of~4-16 kpc (see text for details).
      The red dashed lines delimit the thick disc/bridge/low-$\alpha$ regions used to derive the kinematic properties in Table~\ref{tab:disc}, while the excluded region for this derivation is shaded red.
      Middle panel: two-dimensional number density of star disc particles, as identified by GMM, in the [Mg/Fe] vs. [Fe/H] plane. Green, blue and purple mark the contours containing 90\% of the high-$\alpha$, bridge and low-$\alpha$ disc stars, respectively.  Bottom panel: Mean age distribution of disc stars in the same plane, with the black contour enclosing the region containing 90\% of the star particles in the disc. The last two panels show disc star particles at $z\sim0.1$.
      }
    \label{fig:global_disc}
\end{figure}

\begin{figure*}
    \centering
    \includegraphics[width=0.85\textwidth]{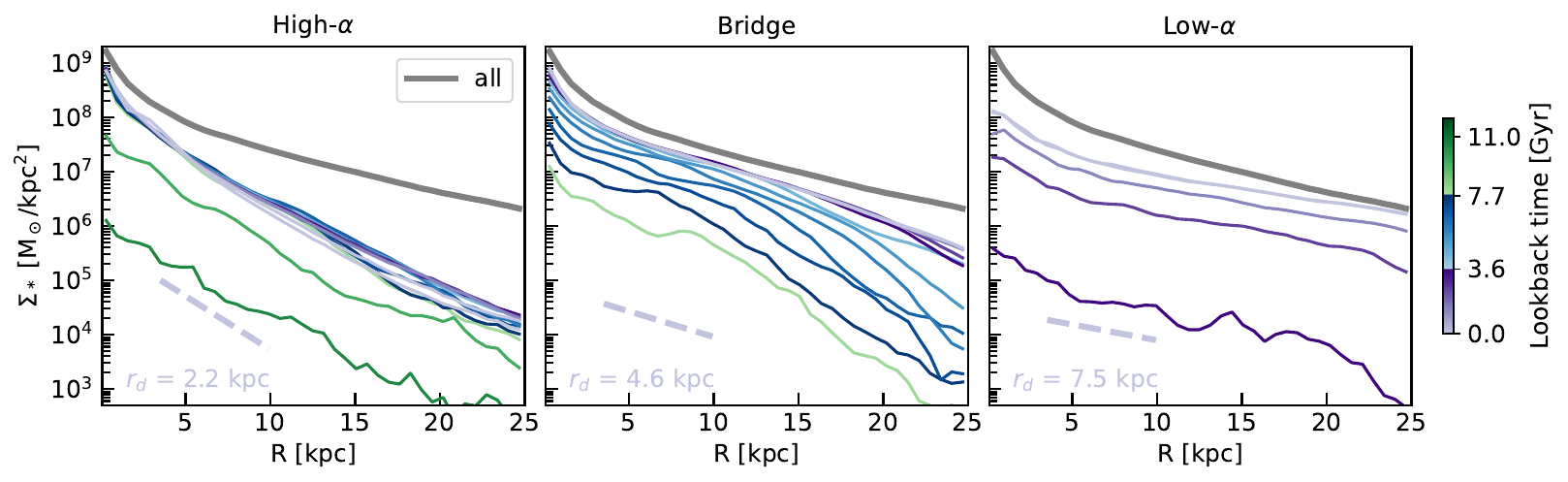}
    \includegraphics[width=0.85\textwidth]{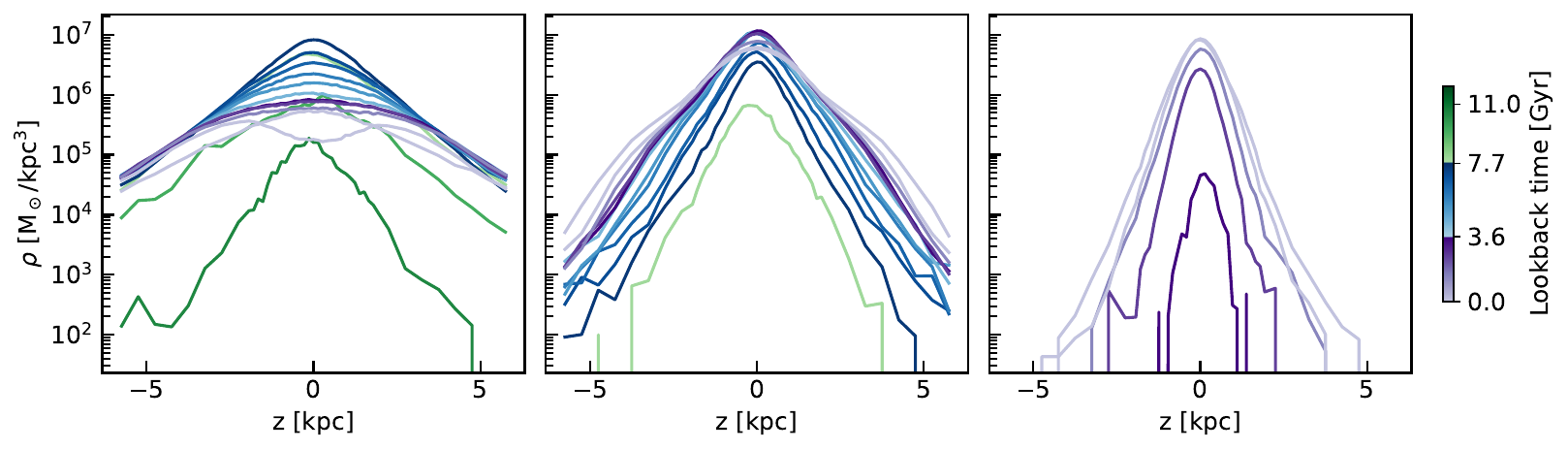}
      \caption{Time evolution of radial stellar surface density (top) and vertical density (bottom) profiles from $z=1.5$ to $z=0$ for the disc components identified using GMM. The thickest solid grey line represents the total disc surface density profile at $z=0$, while the dashed lines show exponential profiles with scale-lengths $r_d$, obtained by fitting the $z=0$ surface density profile of each substructure. The vertical profiles are shown in a cylindrical bin within $r_d\pm1\,\rm kpc$ (see e.g.~\citealt{2021ApJS..254....2P}), where $r_d$ is calculated at each lookback time.
    }\label{fig:disc_time_evol}
\end{figure*}

Interestingly, while the GMM analysis is based solely on ages and kinematics, the identified components also exhibit clear differences in their chemical compositions (Fig.~\ref{fig:global_disc}) and spatial distributions (Fig.~\ref{fig:disc_time_evol}), highlighting their distinct evolutionary histories. Because of the clear location on the [Mg/Fe]-[Fe/H] plane of the three disc components, we name them the ``high-$\alpha$'', ``bridge'' and ``low-$\alpha$'' discs. Furthermore, the three discs match the main disc's formation phases in \textsc{Romeo}: 
\begin{itemize}
\item The high-$\alpha$ disc phase marks the onset of coherent rotation and the settling down in the inner galaxy of a radially compact disc with high velocity dispersion. As shown in Figs.~\ref{fig:global_disc_1Dages} and~\ref{fig:GMM_discs_kinematics}, the spin-up phase timescale is similar to what is expected in the Milky Way, but occurs at younger ages, between 8-11 Gyr. This is consistent with previous analyses of disc formation in \textsc{Romeo} \citep{yu_bursty_2021,yu_born_2023,mccluskey_disc_2024}. Notably, the thick disc formation coincides with an increase in the bursty SFR.
\item As the SFR drops and transitions from bursty to steady at around 7-8 Gyr ago (see bottom panel of Fig.~\ref{fig:general_properties}), consistent with the estimate in \citealt{2023MNRAS.520.1672P}, a thinner intermediate-$\alpha$ and more radially extended disc is able to develop (see bottom central panel of Fig.~\ref{fig:disc_time_evol} compared to the left panel). This disc could correspond to the bridge region identified in the Milky Way~\citep{2021MNRAS.503.2814C}.
\item Finally, about 3.6 Gyr ago (at $z\sim 0.3$), a mild increase of the overall SFR, coupled with a dilution of the gas in the disc as we shall see in Sec.~\ref{subsec:tracks}, enables the onset of the disc bimodality and the growth of the low-$\alpha$ sequence. This timing agrees with that inferred by \citet{2025MNRAS.537.1571P} using a different methodology.
\end{itemize}

The panels of Fig.~\ref{fig:global_disc} show the number density of a sample of giants from Gaia DR3 cross-matched with APOGEE DR17, along with the number density and age distribution of disc stars, as identified by GMM, in the [Mg/Fe]-[Fe/H] plane at $z=0.0998$. The Gaia and APOGEE data preparation and selection criteria is described in App.~\ref{app:GaiaApogee}. 
It can be seen from this figure that the distribution of star particles in {\sc Romeo}, like in the Milky Way, is bimodal in [Mg/Fe]. However, in the Milky Way, the $\alpha$-sequences are broader and more clearly separated along the [Mg/Fe] axis. In particular, in {\sc Romeo}, the difference in the peaks in the bimodal one-dimensional [Mg/Fe] distribution at different galactocentric distance is systematically lower than 0.1 dex, compared to at least 0.2 dex in the MW. This appears to be the case for all the FIRE-2 galaxies in~\cite{2025MNRAS.537.1571P} (see their Fig. A1), which likely points to limitations in the subgrid star formation model used in the simulation, such as the use of supernova metal yield prescriptions that do not account for metallicity dependence. Additionally, this systematic lower difference may reduce the number of simulated galaxies identified as having an $\alpha$-bimodality in~\cite{2025MNRAS.537.1571P}. Finally, another striking difference compared to the Milky Way is the relative prominence of the bridge region in {\sc Romeo} with respect to the analogous counterpart in the MW.

Fig.~\ref{fig:disc_time_evol} illustrates the spatial evolution of each of the identified disc substructures over time.  At $z = 0$, the low-$\alpha$ disc shapes the outer regions of the galaxy, extending beyond $\sim$ 15 kpc, while the bridge dominates the inner regions up to roughly 10 kpc. During the bar epoch, the high-$\alpha$ disc appears to undergo a gradual contraction and its vertical profile deviates from a $\rm sech^2$ profile, consistent with the Milky Way's qualitative trend~\citep{2020A&A...638A..76Q}. Although it is tempting to relate these spatial changes to the unexpected increase in average rotation of the thick disc at later times and the onset of the bar, further analysis, which is beyond the scope of this paper, would be required to reach this conclusion.

\subsection{Radial trends}

\begin{figure*}[h]
    \centering
    \includegraphics[width=\textwidth]{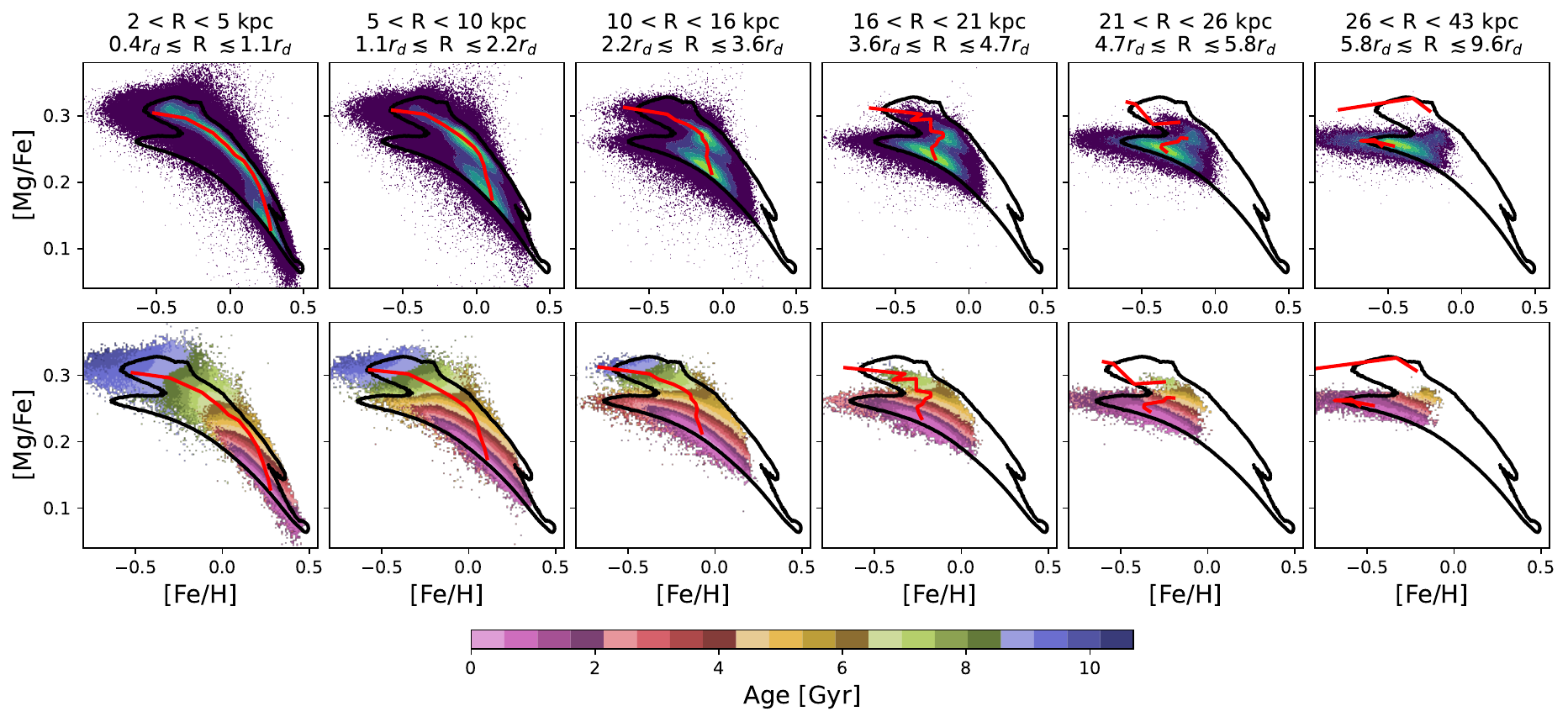}
      \caption{Top: two-dimensional number density of disc star particles within concentric rings with $|z| < 1\,\rm kpc$. The number density distribution is normalised within each ring. Bottom: Mean age distribution of disc stars. The red lines depict the chemical evolutionary track of stars formed from in-situ gas in each of the rings. In the case of the outer rings, the sequence is discontinuous, as there are periods when no stars form from in-situ gas, especially during the spin-up/high-$\alpha$ disc and bridge/intermediate-$\alpha$ phases. The black contours enclose the region containing 90\% of the star particles in the disc.
      }
         \label{fig:global_disc_radial}
\end{figure*}

In order to study the radial trends of the chemical composition of the disc's components, the stellar disc was restricted vertically to $|z| < 1$ kpc and binned into concentric annular cylinders. The bin sizes were chosen to facilitate comparison with the radial structure observed in the MW at similar multiples of the disc scale-length $r_d$ (see Fig. 8 in~\citealt{2023ApJ...954..124I}). For the MW, we adopt a scale-length of 2.6 kpc, while for {\sc Romeo}, a scale-length of 4.5 kpc was determined by fitting the entire disc surface density profile.  
The top panel of Fig.~\ref{fig:global_disc_radial} shows two-dimensional distributions [Mg/Fe] versus [Fe/H] of star disc particles at $z=0.0998$ within each ring. Although see Fig.~\ref{fig:global_disc_radial_components} for the two-dimensional number density of each GMM disc component. 

Overall, the radial structure of the chemical bimodality in {\sc Romeo} resembles that of the MW~\citep{2015ApJ...808..132H, 2020A&A...638A..76Q, 2023ApJ...954..124I}, with stars in the low-$\alpha$ disc becoming progressively more metal-poor at larger galactocentric distances. We also find that the ratio of stars with high-$\alpha$ to those with low-$\alpha$ increases with higher $|z|$. This supports the conclusions of~\cite{2026MNRAS.545f1551O} that this vertical structure is likely a general result of disc formation in MW-mass galaxies. There are, nonetheless, two very notable differences compared to the MW. Firstly, the high and intermediate-$\alpha$ discs in \textsc{Romeo} remains as dominant as (or even more than) the low-$\alpha$ component out to $3.6\times r_d$ ($\sim 16$ kpc), unlike in the MW. \textsc{Romeo} ---with its more extended stellar disc--- still hosts a notable high-$\alpha$ plus bridge component out to almost 4.7$\times r_d$ ($\sim 20$ kpc), whereas in our Galaxy, the high-$\alpha$ population drops sharply beyond 9 kpc or 3.6$\times r_d$. This may suggest that after the spin-up phase and formation of the thick disc, the MW experienced a drop in the SFR, leading to an underpopulated bridge. 

Secondly, the change in the location of the [Mg/Fe] vs [Fe/H] plane of the low-$\alpha$ sequence as a function of $R$ is more pronounced in {\sc Romeo} than in the Milky Way. This second difference could be attributed to subdominant radial stellar migration during the development of the low-$\alpha$ in {\sc Romeo} (see App.~\ref{app:radial_migration}) or may point to overly large radial fluxes, which could be alleviated by the presence of cosmic ray physics~\citep{2022MNRAS.509.4149T}, not implemented in the current galactic model. Both processes would dilute metallicity differences between radial bins.

The bottom panel of Fig.~\ref{fig:global_disc_radial} depicts the age distribution of the star disc particles. By comparing the top and bottom panels of this figure, it can be inferred that there is a deficit of stellar particles with ages around 3-4 Gyr (as can also be seen in Figs.~\ref{fig:global_disc_1Dages} and \ref{fig:global_disc}). This deficit driven by a dropping SFR (see bottom panel of Fig.~\ref{fig:general_properties}) is one of the necessary conditions for forming two separate $\alpha$-sequences in the absence of radial migration. Following this dip in the number density of stars, the low-$\alpha$ population begins to dominate over the intermediate-$\alpha$ sequence. In our Solar neighbourhood, a dearth of stars was found at ages around 6 Gyr~\citep{2020A&A...640A..81N}, although \cite{2025A&A...704A.258F} instead found a shortage at 3 Gyr. It should be noted that this shortage of stars in the Milky Way is not associated with the onset of the low-$\alpha$ disc, which is estimated to have occurred earlier, around 8-10 Gyr ago~\citep{2025A&A...704A.258F}.

\subsection{Gas and stellar chemical tracks}
\label{subsec:tracks}
\begin{figure}
    \centering
    \includegraphics[width=0.75\columnwidth]{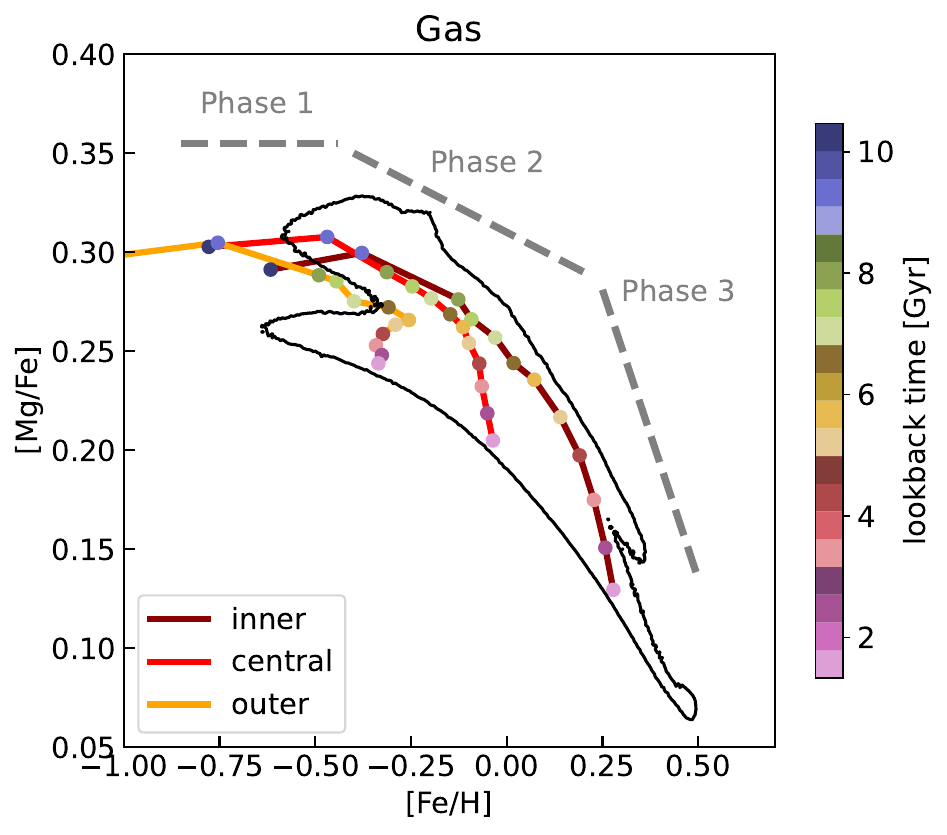}
    \includegraphics[width=\columnwidth]{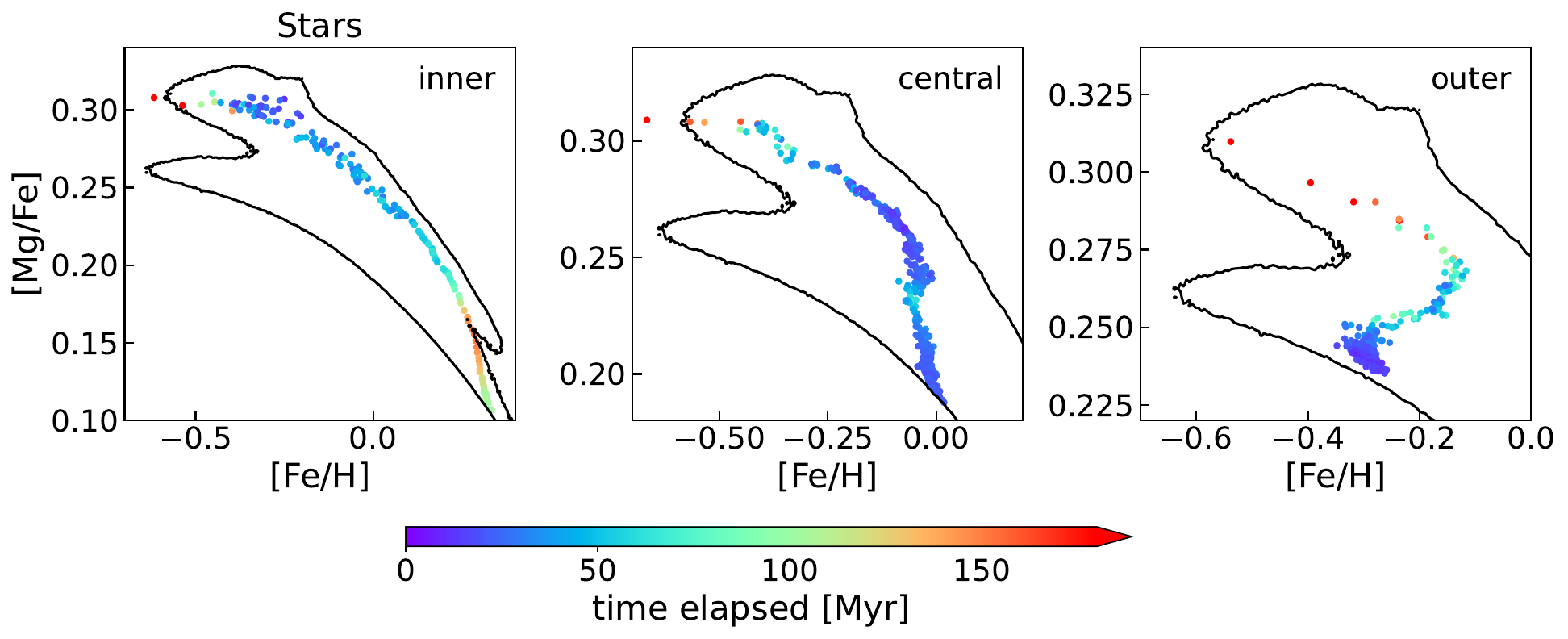}
      \caption{Top panel: gas chemical tracks in the inner, central, and outer disc regions. Each track is constructed by, at each snapshot, selecting the gas within the corresponding region and computing its (mass weighted) average chemical properties. Grey dashed lines indicate schematically the three general trends exhibited by the chemical tracks. 
      Bottom panels: dynamically-evolved stellar chemical tracks for the inner (left), central (middle), and outer (right) regions. These are built by selecting, at the final snapshot, the stars located in each region and tracing back their formation times and birth chemistry. Stars are plotted within the track of the region in which they reside at the end, regardless of their place of birth. 
      Each point represents $10^3$ stars, coloured by the time required to form this number of stars; shorter/bluer (longer/redder) formation times correspond to periods of higher (lower) SFR, thus tracing the SFH of each region.
      }
    \label{fig:chemical_tracks}
\end{figure}

Figure~\ref{fig:chemical_tracks} shows the chemical tracks for the gas (left panel) and for the stars (right panels) in three concentric annular rings. Namely, in an inner, central and outer ring spanning $0 < R\,[\rm kpc] < 5$, $5 < R\,[\rm kpc] < 16$ and $16 < R\,[\rm kpc] < 26$, respectively. Each cylindrical annuli is restricted vertically to $|z| < 2$ kpc. While stellar tracks of each annular ring are built by selecting the stars that at the end of the simulation are within that bin, regardless of their birth location, gas tracks are constructed by, at each snapshot, accounting for the gas within a given annular cylinder. The dynamically-evolved stellar tracks closely resemble that of the gas. Therefore, stars largely preserve their birth galactocentric distance and stellar migration in Romeo is, on average, negligible or subdominant, as can also be seen in App.~\ref{app:radial_migration}. As stars retain the chemical conditions of the cold gas-phase at their birthplace and time of formations, in the absence of substantial radial migration, gas chemical evolution in a given disc region dictates the chemical evolution of the stars.

We can identify three phases in the chemical evolution of the gas and dynamically-evolved stellar tracks, based on changes in the slope of these tracks. Interestingly, these three phases are associated with the three disc components identified by GMM and the three phases of the SFH. Namely:
\begin{itemize}
    \item The stars in the thick disc are characterised by high-$\alpha$ abundances with relatively constant values of [Mg/Fe] with increasing [Fe/H] values. These stars form during an early epoch with a rising and bursty SF, when core-collapse supernovae (SNe) exceed SNe type Ia.
    \item Next, the star formation begins to decrease and enters a smoother phase. During this period, stars form in the bridge region with intermediate-$\alpha$ values. Their [Mg/Fe] abundance ratios decrease progressively with higher values of [Fe/H], as SNe type Ia begin to outnumber core-collapse SNe, thus dominating the mass yields of Fe mass over the production of Mg. Before the end of bridge formation, around 5 Gyr ago, chemical tracks show a steepening of the slope and even a decrease in [Fe/H] in the outer disc region. As we shall see in Section~\ref{subsec:corona}, this dilution of the metal content in the disc is caused by the hot, coronal gas.
    \item Finally, a mild increase in the overall SFR, together with the metallicity dilution, enables the emergence of a sequence of disc stars with low-$\alpha$ abundance ratios. 
\end{itemize}
It should be noted that, in the absence of stellar migration as in {\sc Romeo}, gas dilution in the disc is a necessary but not sufficient condition for generating the separation and formation of two distinct $\alpha$ sequences. The other necessary condition is either a decrease in the SFR or a rapid and sudden decrease in the [Mg/Fe] abundance ratio of the gas, so that a deficit occurs in the number of stars in the region between the two sequences. In the case of {\sc Romeo}, the first scenario applies, as can be seen in the lower panels of Fig.~\ref{fig:chemical_tracks} by the slight increase in the time elapsed to form $10^3$ stars (decrease in SFR) in the region separating the high and low-$\alpha$ sequences and its subsequent moderate decrease (increase in SFR).

\begin{figure}
    \centering
    \includegraphics[width=0.5\columnwidth]{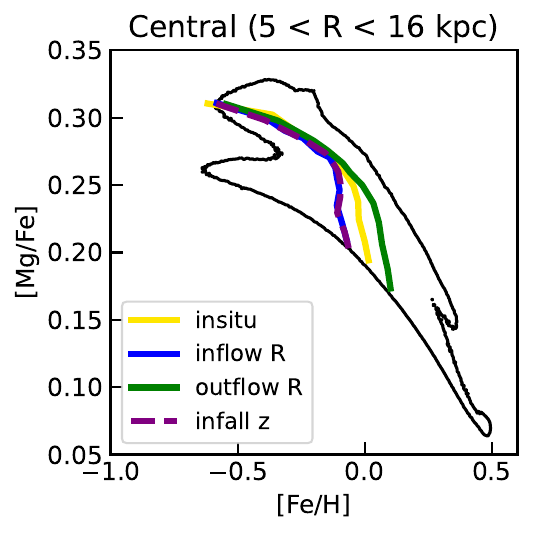}
      \caption{Chemical evolutionary tracks of stars, separated according to whether they formed from in-situ gas or from infalling gas arriving from different directions.
      }
    \label{fig:gas_into_stars_track}
\end{figure}

Finally, Fig.~\ref{fig:gas_into_stars_track} shows the chemical tracks of stars in the central bin. In here, tracks are constructed in such a way that the stars are grouped according to whether they formed from the in-situ gas or from gas falling into the bin from different directions. It is clear from this plot that radial outflows, primarily driven by stellar winds from massive stars and SNe explosions, carry more metal-rich material outward, producing, on average, more metal-rich tracks. On the other hand, radial inflows bring in lower metallicity gas. Despite some mixing between the infalling and the in-situ gas\footnote{Gas tracks when separated by their origin show greater dispersion in metallicity compared to stellar tracks.}, stars of the same age at a given radius show significant dispersion in metallicity, reflecting the chemically diverse origins of their progenitor gas. This implies that radial stellar migration is not necessary to explain the chemical spread in a stellar disc, and that the commonly adopted assumption of instantaneous mixing in chemical analytical models cannot explain the observed dispersions in metallicity and [Mg/Fe] observed in the {\sc Romeo} simulated galaxy. 

\section{Physical drivers of disc evolution}
\label{sec:chemical-evol}

\subsection{Gas flow rates}
\label{subsec:flow_rates}
\begin{figure}
    \centering
    \includegraphics[width=0.6\columnwidth]{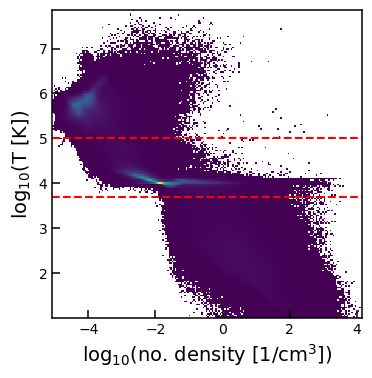}
      \caption{Temperature versus number density distribution of the gas in {\sc Romeo} at $\rm t_{lb}=5.16\, Gyr$. The horizontal lines mark the cuts used to define the cold, warm and hot phases, which separate the gas into the cold disc, the cooling bridge connecting the disc with the corona, and the hot corona.
      }
    \label{fig:gas_phases}
\end{figure}

We calculate gas inflow and outflow rates across the boundaries of the three annular rings defined in Sect.~\ref{subsec:tracks} and as described in App.~\ref{app:gas_flow_rates}. In this way, we evaluate the role of vertical gas accretion and radial gas flows in the development of the three phases of the chemical evolution of the stellar disc. Fluxes are separated by gas temperature, which helps identify their origin. In particular, we distinguish three gas phases: cold ($\rm T\,[K] < 5\times10^3$), warm ($5\times10^3 < \rm T\,[K] < 10^5$), and hot ($\rm T\,[K] > 10^5$) gas.
As can be seen in Fig.~\ref{fig:gas_phases}, these cuts separate the gas into the hot, low-density corona, whose gas gradually cools and shifts into the transition region at intermediate densities, from where it continues to lose energy and slowly accretes onto the cold, dense disc. This transition region also contains originally cold gas from gas-rich mergers that was stripped by ram pressure and heated as its bulk kinetic energy is dissipated into thermal energy.

The above temperature boundaries differ slightly from those adopted by \cite{2023MNRAS.524.4091B}. The lower cold–warm threshold allows us, on the one hand, to identify as disc gas only the cold component confined to the disc mid-plane and, on the other hand, to account for differences between the radiative feedback model used in \cite{2023MNRAS.524.4091B} and that adopted in {\sc Romeo}. In particular, because FIRE-2 does not impose a characteristic temperature for photoionised gas, the slightly lower threshold adopted here ensures that the intermediate-temperature cooling-bridge phase fully encompasses gas at densities $\sim10^{-2}\,\rm{cm^{-3}}$, whose origin is predominantly the hot corona (see Fig.~\ref{fig:gas_phases}). Additionally, we have verified that adopting either our warm–hot boundary or that of \cite{2023MNRAS.524.4091B} does not affect the results of our analysis.

\begin{figure*}
    \centering
    \includegraphics[width=0.8\textwidth]{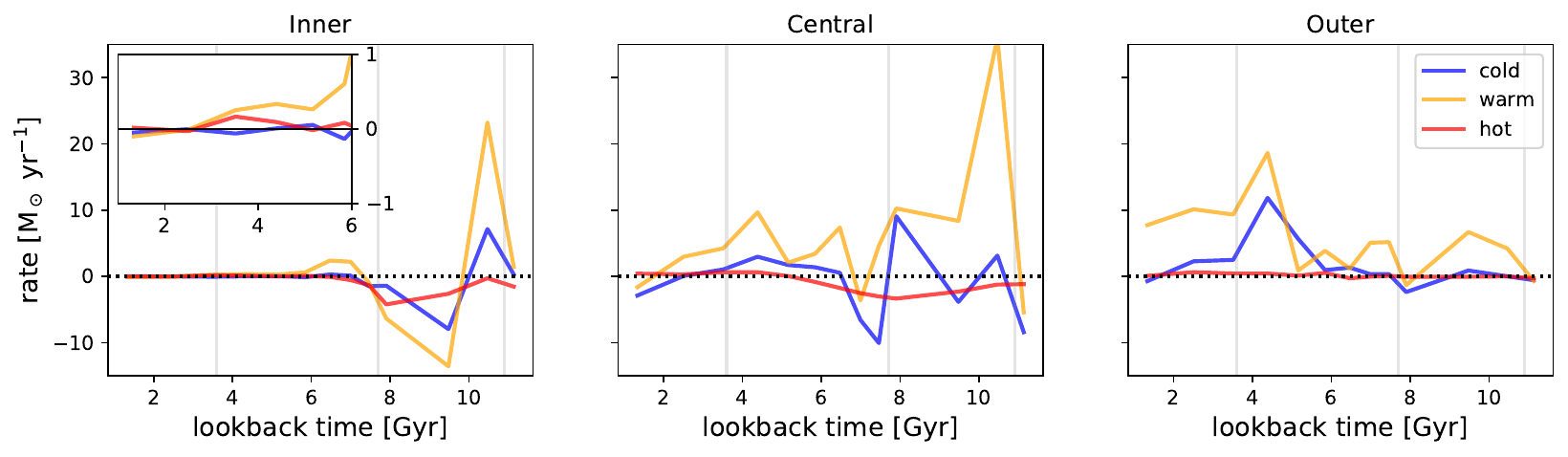}
    \includegraphics[width=0.8\textwidth]{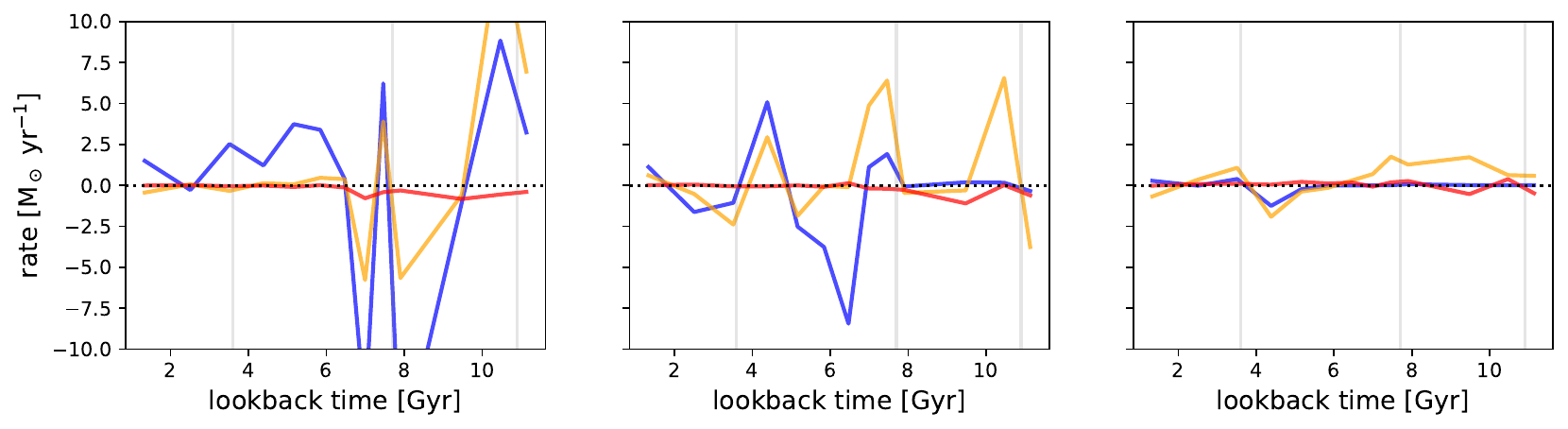}
      \caption{Vertical (top panels) and radial (bottom panels) net (inflow minus outflow) gas flow rates through an inner ($0 < R\,[\rm kpc] < 5$), central ($5 < R\,[\rm kpc] < 16$) and outer ($16 < R\,[\rm kpc] < 26$) annular rings for cold ($\rm T\,[K] < 5\times10^3$), warm ($5\times10^3 < \rm T\,[K] < 10^5$, middle) and hot ($\rm T\,[K] > 10^5$) gas. 
      Positive net fluxes correspond to inflow (i.e., gas entering the corresponding ring), while negative values indicate gas outflow.
      The vertical lines in the panels indicate the lookback times (3.6, 7.7 and 10.9 Gyr ago) that separate the formation and evolution of the three disc phases identified in {\sc Romeo}.
      }
    \label{fig:flow_rates}
\end{figure*}

Gas can enter the disc through accretion from the galactic corona, via gas-rich mergers, or through galactic fountains, where stellar feedback ejects gas from the disc into the corona before it eventually falls back \citep{2008MNRAS.386..935F, 2019MNRAS.489.4233M, 2023MNRAS.524.4091B}. The typical fountain cycle is $\sim 100\,\rm Myr$ \citep{2008MNRAS.386..935F}, so outflow and inflow rates are expected to track each other with a delay of this order, although, on average, these rates balance each other out. Such a temporal shift cannot be seen in Fig.~\ref{fig:flow_rates} as the time span between available snapshots is much larger. Coronal accretion contributes primarily to the warm phase, mergers supply both cold and warm gas, and fountains predominantly recycle cold material. 

The upper panels of Fig.~\ref{fig:flow_rates} showing the vertical net (inflow minus outflow) rates, indicate that, along the disc formation, the vast majority of the gas vertically accreted onto the disc is warm, as can be seen by the overall positive net rate of this gas phase, especially in the central and outer disc regions. This offset is not as substantial as in \cite{2023MNRAS.524.4091B}, likely reflecting differences in the gas physics model and the absence in {\sc Romeo} of positive feedback from the efficient mixing between gas ejected from the disc and the hot metal-poor corona, which would otherwise accelerate cooling and boost star formation~\citep{2023MNRAS.524.4091B}. 
The accreted warm gas is mainly hot coronal gas that cools down, thus producing an overall positive offset of warm inflows relative to warm outflows. We also arrived to the conclusion that hot coronal gas is the main source of accreted gas by first constructing SUBFIND halo catalogues \citep{2001MNRAS.328..726S} at two lookback times selected just after and before the onset of the thick disc, specifically 9.5 Gyr and 11.6 Gyr ago. We then identified the gas cells that accreted vertically in the disc and located the group or subhaloes to which each of these cells belonged in the catalogues. The largest percentage of cells belonged to the ``outer fuzz'', which we identified as the intergalactic medium that slowly accretes onto the halo corona and from there into the disc.  

\begin{figure}
    \centering
    \includegraphics[width=\columnwidth]{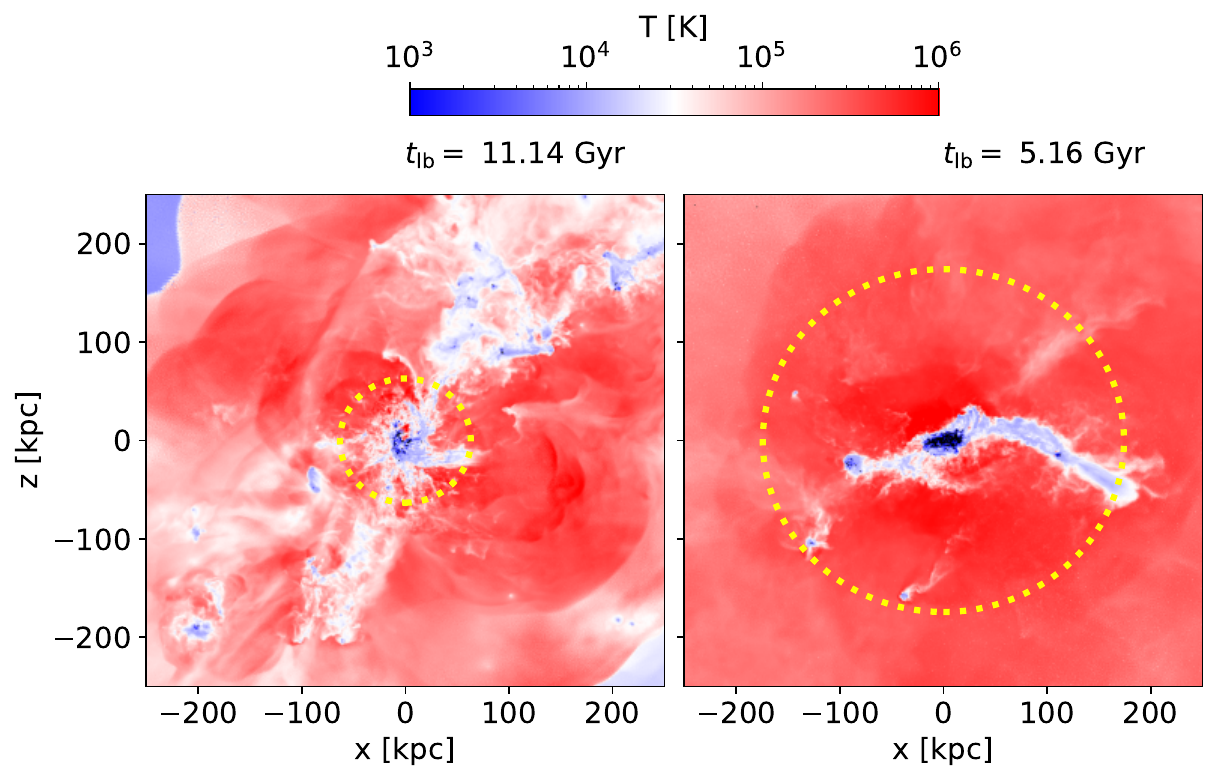}
      \caption{Gas temperature at different lookback times. These are number density-weighted profiles along the perpendicular direction, enhancing the visibility of underdense regions by reducing the dominance of high-density areas. The reference frames are aligned with the principal axes of {\sc Romeo} at the last simulated snapshot. The yellow circle marks the virial radius $R_{200, c}$ at each time.
      }
    \label{fig:gas_portrait}
\end{figure}

Two other trends can be identified from the accretion rates. First, an irregular and accentuated profile of the net vertical rate of warm and cold gas, with high peaks of gas inflow and steep descents where outflows dominate. This initial period, which lasted until approximately 7.7 Gyr ago (corresponding to the beginning of the formation of the bridge or intermediate-$\alpha$ disc), gives rise to a bursty star formation, and is associated with multiple gas-rich mergers that results in the formation of a compact thick disc exhibiting high velocity dispersion like in \cite{2004ApJ...612..894B}. This initial chaotic epoch of hierarchical clustering can be seen in the left panel of Fig.~\ref{fig:gas_portrait} that displays the galaxy right before the thick disc starts to settle. At this early time of the galaxy's collapse, the gas number density traces the primitive filamentary structure of the dark matter Cosmic Web~\citep{1978IAUS...79..241J, 1980Natur.283...47E} and gas is accreted from the intergalactic medium through a filament, with cold gas clouds flowing through it.  

After this chaotic epoch, the vertical inflow and outflow rates of the cold phase nearly balance, consistent with recycling-dominated flows, except for a peak around 4-5 Gyr ago, which coincides with a peak in the warm phase. This peak is attributable to minor mergers with gas-rich satellites, as can be seen in the right panel of Fig.~\ref{fig:gas_portrait}. Specifically, this panel shows the gas temperature at a lookback time close to the pericentric passages of two minor mergers and clearly reveals cold gas filaments connecting the outer edges of the disc in {\sc Romeo} with the satellites. The accretion of this gas produces the slight increase in the overall SFR of {\sc Romeo} which enables the development of the low-$\alpha$ sequence.

Additionally, the bottom panels of Fig.~\ref{fig:flow_rates} shows the radial net (inflow minus outflow) rates. From here, we would like to highlight that, differently from the vertical accretion, radial accretion is dominated by cold flows in the inner and central disc regions. The inflow of cold, metal-poor gas is mainly a dynamical consequence of low-angular momentum gas accreting vertically onto the rotating disc. Since this gas has lower specific angular momentum than the disc material, part of it must flow radially inward to conserve angular momentum~\citep{1981A&A....98....1M}. Thus, although late gas accretion mainly occurs in the outer disc, some of this cold, metal-poor gas provides new material to fuel the star formation of the low-$\alpha$ sequence in the inner regions of the disc.

\subsection{Galactic corona}
\label{subsec:corona}

\begin{figure}
    \centering
    \includegraphics[width=\columnwidth]{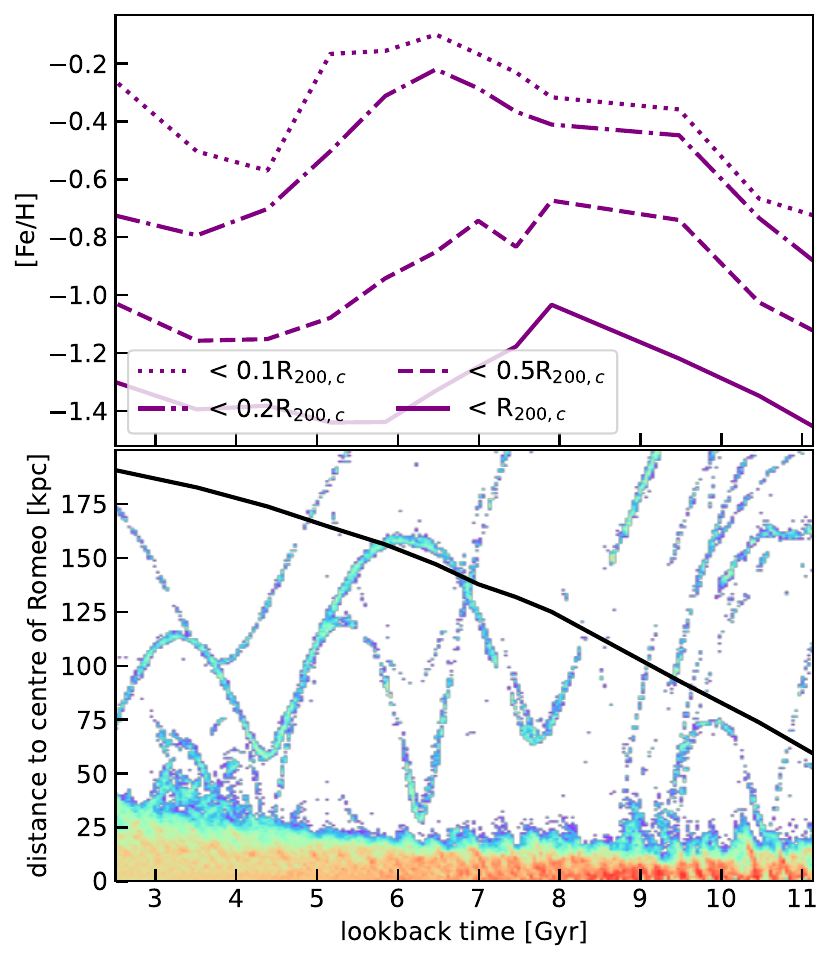}
      \caption{Top: time evolution of the mass-weighted [Fe/H] of the gaseous corona as a function of lookback time, with linestyles indicating the average metallicity within successive radial cuts (gas within $|z|<2$ kpc was excluded to avoid contamination from the disc): $r \leq 0.1R_{200, c}$, $r \leq 0.2R_{200, c}$, $r \leq 0.5R_{200, c}$, and the entire corona.
      Bottom: density of forming stars as a function of formation time (x-axis) and galactocentric radius $r$ (y-axis), with red indicating higher and blue lower densities. This highlights both the inside-out growth of the disc and the orbits of gas-rich merger events. The black line traces the virial radius $R_{200, c}$ as a function of lookback time.
    }\label{fig:corona_fe}
\end{figure}

Figure~\ref{fig:corona_fe} illustrates how the metallicity of the hot corona evolves over time in connection with the accretion history of {\sc Romeo}. In the top panel, one can observe a metallicity dilution in the outer layers of the corona around 8 Gyr ago. The onset of this dilution in the corona appears to be linked to the infall of a minor merger crossing the virial radius of {\sc Romeo} (see bottom panel). In fact, in the snapshot just after this infall ($t_{\rm lb}=7.9\,\rm Gyr$), we separated the coronal gas to whether it belonged right before the infall at $t_{\rm lb}=9.5\,\rm Gyr$ to the ``outer fuzz''\footnote{The ``outer fuzz'' corresponds to the set of particles not associated with any Friends-of-Friends halo constituting a diffuse, unbound component surrounding bound structures~\citep{2001MNRAS.328..726S}. This component is therefore identified as the intergalactic medium.}, other halos or {\sc Romeo}. The dilution is only found in the gas belonging to other halos. Although this component accounts for only $\sim 25\%$ of the total coronal gas mass (with the minor merger contributing roughly 15\% and the ``outer fuzz'', contributing about $70\%$), it is sufficient to drive the observed dilution.
At $\rm t_{lb}=11.6\, Gyr$ ($z=3$), the minor merger contains slightly less than $1\%$ of {\sc Romeo}'s stellar mass, about $6\%$ of its gas mass, and roughly $6\%$ of its total mass. 

As can be seen from the top panel of the same figure, the dilution propagates from the outer to the inner layers of the galactic corona. In the inner regions, the metallicity decline happens around 5 Gyr ago, coinciding with the halt --and even reversal-- of the [Fe/H] increase in the gas of the disc, as shown by the chemical tracks in the upper panel of Fig.~\ref{fig:chemical_tracks}. A comparable decline for gas accreted within 10 kpc (similar to $0.1 R_{200, c}$) was also reported by \citet{2025MNRAS.537.1571P} (see their Fig. 9), thus indicating consistency in the different analyses.
These results support the interpretation that accreted metal-poor gas from a gas-rich minor merger drives the decline, first in the corona and subsequently in the disc of {\sc Romeo}.

The bottom panel of this figure also shows the pericentric passages of the gas-rich minor mergers, around 4-5 Gyr ago, responsible for the increase in the inflow rates of the cold and warm gas-phases in the central and outer disc regions (see also right panel of Fig.~\ref{fig:gas_portrait}). The accretion of gas from these mergers results in a mild increase in the overall SFR and the onset of the low-$\alpha$ sequence. In addition, we can observe that following the pericentric passages, the disc expands, i.e., the accretion of gas from these close approaches facilitates the inside-out growth of the disc. This is consistent with the trends shown in~\cite{2025MNRAS.537.1571P}, where the FIRE-2 simulated galaxies with strongest bimodalities exhibit more extended discs.  

\section{Discussion}
\label{sec:conclusions}

\subsection{Comparison with previous work}

We observe a clear [Mg/Fe]-[Fe/H] dichotomy in the {\sc Romeo} simulated galaxy, an example of a system with no major mergers and with no bar at present. It should be noted that the high and low-$\alpha$ sequences are, nonetheless, less separated in [Mg/Fe] than observed in the Milky Way. A similar weak separation is seen in other FIRE-2 galaxies analysed in~\cite{2025MNRAS.537.1571P}, which may point to limitations in the subgrid star formation model, such as the metal yield prescriptions adopted. 

We note that while the absence of certain physical processes in {\sc Romeo}, such as the lack of active galactic nuclei feedback, may affect the generality of our findings, our results indicate that neither stellar radial migration nor major mergers are required for a galaxy to develop a disc's bimodal chemical structure. Instead, we argue that $\alpha$-bimodalities seem to be a natural consequence of inside-out disc growth in a hierarchical formation scenario. This is consistent with trends shown in~\cite{2025MNRAS.537.1571P}, where 11 FIRE-2 simulated galaxies are analysed (including {\sc Romeo}) and those with a strongest bimodality have the most extended stellar discs. 

In the scenario proposed here, the high-$\alpha$ and low-$\alpha$ discs form in two distinct star-formation modes, consistent with previous studies \citep{2019MNRAS.484.3476C,2021MNRAS.501.5176K}. Simmilar to those works, stars on the high-$\alpha$ sequence form, on average, in regions of higher SFR density. Although early star formation is spatially more patchy, whereas at later times it is smoother along the disc, we do not find spatially extended clumps of high SFR density comparable to those reported by \cite{2019MNRAS.484.3476C}. While \citet{2021MNRAS.501.5176K} showed that bimodality can arise naturally from inside-out disc growth in isolation, in {\sc Romeo} minor mergers are required: first to dilute the galactic corona and subsequently to supply fresh gas that fuels the low-$\alpha$ sequence.

\subsection{Conclusions}
We find that rather than a strict chemical dichotomy of chemical thick versus thin disc, \textsc{Romeo} exhibits a trichotomy with three main disc assembly phases: high-$\alpha$, bridge and low-$\alpha$ discs, each with distinct formation times, chemical tracks, and kinematic signatures across galactocentric radii. The development of these three disc phases is as follows:
\begin{itemize}
\item Similar to the (semi)isolated galaxy of~\cite{2004ApJ...612..894B}, the compact, high-$\alpha$ thick disc develops during the spin-up phase, characterised by a turbulent period of hierarchical clustering and a high, bursty star formation rate, resulting in a disc with high velocity dispersion. The spin-up phase, comparable to that of the Milky Way~\citep{2022MNRAS.514..689B}, is short and lasts $\sim$2 Gyr. However, in Romeo it occurs somewhat later and at higher metallicities, between 8-11 Gyr ago, compared to 10-13 Gyr ago in the Milky Way. We would like to note that the disc in {\sc Romeo} is the first to settle among the Milky Way-mass galaxies in the FIRE-2 suite~\citep{mccluskey_disc_2024}. This aligns with results from other hydrodynamical simulations that tend to form discs later than observed in the Milky Way~\citep{2022MNRAS.517..832I, 2024ApJ...962...84S, 2025arXiv250611840M}. 
\item Subsequently, a thinner, more extended disc grows as star formation starts decreasing and transitions into a more steady phase, in agreement with early findings by \cite{1976MNRAS.176...31L}. This disc may correspond to the bridge region identified in the MW~\citep{2021MNRAS.503.2814C}, though the prominence of this feature in the MW is significantly weaker than in {\sc Romeo}. Recent analyses of the MW SFH indicate a quenching phase around 9-10 Gyr ago~\citep{2025A&A...704A.258F}, coinciding with the epoch of bridge formation~\citep{2021MNRAS.503.2814C}. This would therefore explain the low populated bridge in the MW, which may be a direct consequence of quenching due to the GSE merger event. 
\item Finally, around 5 Gyr ago, gas accretes onto the disc due to pericentric passages of gas-rich minor mergers, causing a mild increase in the SFR and favouring the inside-out growth of the stellar disc.
This, together with the halting of the evolution of the metals due to dilution driven by the galactic corona, fuels the development of a low-$\alpha$ disc. Although late gas accretion mainly occurs in the outer disc, as expected in an inside-out growth scenario, some of this gas radially flows inwards providing new material to fuel star formation in the inner disc.
\end{itemize}

\noindent To sum up, our conclusions are as follows:
\begin{itemize}
    \item Stellar discs can develop an $\alpha$-bimodality as a natural consequence of inside-out disc growth driven by the hierarchical assembly of gas-rich minor mergers. This is not necessary in conflict with observations, which suggest that minor mergers contribute little to the gas accretion budget in the local Universe~\citep{2014A&A...567A..68D}. 
    \item As a testable prediction of this scenario, we expect that $\alpha$-bimodalities are common among Milky Way–mass disc galaxies and do not correlate strongly with either the presence of a bar or the galaxy’s position within the Cosmic Web. Moreover, since the high- and low-$\alpha$ sequences form through slow secular processes, stars in both components are expected to be close to dynamical equilibrium, and the mean of the metallicity distribution function of stars in the low-$\alpha$ sequences at a given radii is constant or only mildly increasing with age.
    Though a proper statistical analysis, beyond the scope of this paper, is required to confirm these hypotheses.
    \item The chemical evolution of the stellar disc cannot be understood without taking into account its interaction with the surrounding galactic corona, which constitutes the main gas reservoir for the disc and is itself replenished mainly through accretion from the intergalactic medium. 
    \item In {\sc Romeo}, at the end of the high-$\alpha$ disc formation, a minor gas-rich merger causes a metallicity dilution of the coronal gas. This dilution propagates from the outer to the inner coronal layers and then to the disc, providing one of the two conditions (in the absence of substantial radial migration) necessary for the formation of two distinct $\alpha$-sequences: the halting (or even the decrease) of metallicity. 
    \item At the same time, the SFR begins to decrease, generating the second necessary condition for the development of a bimodality: a dip in the number density of stars, followed by gas accretion at the pericentric passages of minor mergers that fuels the onset of the low-$\alpha$ sequence.
    \item Finally, we find that instantaneous mixing does not hold in {\sc Romeo} and the spreads in the stellar [Fe/H] and [Mg/Fe] abundance ratios at a given galactocentric distance reflect the diverse origins of infalling gas, highlighting the importance of accounting for both vertical and radial gas flows.
\end{itemize}
Looking ahead, determining the ubiquity of $\alpha$-bimodalities (see e.g. \citealt{2024A&A...691A..61P}) and their correlation with galaxy’s position in the Cosmic Web is crucial, as large-scale environment may regulate gas content, inflow timing and merger history, influencing disc chemistry and the emergence of multiple chemical sequences (e.g.~\citealt{2023A&A...677A..91K}).

\begin{acknowledgements}
We send a special thanks to Claudio Dalla Vecchia for providing the necessary scripts to construct SUBFIND halo catalogues along with useful discussion. We also thank Jenna Samuel and Andrew Wetzel for valuable discussions. 

MB is funded by the European Union (ERA Fellowship, DiaLoGues, 101180670). This work was supported by the Estonian Research Council grants (PSG938, PRG1006, PRG3034) the Estonian Ministry of Education and Research (grant TK202) and the European Union's Horizon Europe research and innovation programmes (EXCOSM, grant No. 101159513, and EXOHOST, grant No. 101079231). 
GB and EFA acknowledge support from the Agencia Estatal de
Investigación del Ministerio de Ciencia, Innovación y Universidades
(MCIU/AEI) under grant "EN LA FRONTERA DE LA ARQUEOLOGÍA GALÁCTICA:
EVOLUCIÓN DE LA MATERIA LUMINOSA Y OSCURA DE LA VÍA LÁCTEA Y LAS
GALAXIAS ENANAS DEL GRUPO LOCAL EN LA ERA
DE GAIA. (FOGALERA)", the European Regional Development Fund (ERDF)
with reference PID2023-150319NB-C21 and PID2023-150319NB-C22. EFA also acknowledges support from HORIZON TMA MSCA Postdoctoral Fellowships Project TEMPOS, number 101066193, call HORIZON-MSCA-2021-PF-01, by the European Research Executive Agency.
HR acknowledges funding from UK Research and Innovation (UKRI) under the UK government’s Horizon Europe funding guarantee (grant No. 10051045). SCB acknowledges the support of the Agencia Estatal de Investigación del Ministerio de Ciencia e Innovación (MCIN/AEI/10.13039/501100011033) under grant nos. PID2021-128131NB-I00 and CNS2022-135482 and the European Regional Development Fund (ERDF) `A way of making Europe' and the `NextGenerationEU/PRTR'. 

The results and figures presented in this work were made possible thanks to the following software libraries: Matplotlib \citep{matplotlib}, NumPy \citep{numpy}, SciPy \citep{2020SciPy-NMeth}, Jupyter \citep{jupyter}, Scikit-learn \citep{pedregosa_scikit-learn_2011}, Gaia-tools (\href{https://github.com/HEP-KBFI/gaia-tools}{https://github.com/HEP-KBFI/gaia-tools}) and GizmoAnalysis (\href{http://ascl.net/2002.015}{http://ascl.net/2002.015}). This last repository was first used in~\cite{2016ApJ...827L..23W}.
\end{acknowledgements}

\bibliographystyle{aa}
\bibliography{references}

@ARTICLE{2023A&A...677A..91K,
       author = {{Khoperskov}, Sergey and {Minchev}, Ivan and {Libeskind}, Noam and {Belokurov}, Vasily and {Steinmetz}, Matthias and {Gomez}, Facundo A. and {Grand}, Robert J.~J. and {Hoffman}, Yehuda and {Knebe}, Alexander and {Sorce}, Jenny G. and {Spaare}, Martin and {Tempel}, Elmo and {Vogelsberger}, Mark},
        title = "{The stellar halo in Local Group Hestia simulations. III. Chemical abundance relations for accreted and in situ stars}",
      journal = {\aap},
         year = 2023,
        month = sep,
       volume = {677},
          eid = {A91},
        pages = {A91},
          doi = {10.1051/0004-6361/202244234},
archivePrefix = {arXiv},
       eprint = {2206.05491},
 primaryClass = {astro-ph.GA},
       adsurl = {https://ui.adsabs.harvard.edu/abs/2023A&A...677A..91K}
}

@article{wetzel_public_2023,
	title = {Public data release of the {FIRE}-2 cosmological zoom-in simulations of galaxy formation},
	volume = {265},
	issn = {0067-0049, 1538-4365},
	url = {http://arxiv.org/abs/2202.06969},
	doi = {10.3847/1538-4365/acb99a},
	language = {en},
	number = {2},
	urldate = {2023-10-23},
	journal = {ApJS},
	author = {Wetzel, Andrew and Hayward, Christopher C. and Sanderson, Robyn E. and Ma, Xiangcheng and Angles-Alcazar, Daniel and Feldmann, Robert and Chan, T. K. and El-Badry, Kareem and Wheeler, Coral and Garrison-Kimmel, Shea and Nikakhtar, Farnik and Panithanpaisal, Nondh and Arora, Arpit and Gurvich, Alexander B. and Samuel, Jenna and Sameie, Omid and Pandya, Viraj and Hafen, Zachary and Hummels, Cameron and Loebman, Sarah and Boylan-Kolchin, Michael and Bullock, James S. and Faucher-Giguere, Claude-Andre and Keres, Dusan and Quataert, Eliot and Hopkins, Philip F.},
	month = apr,
	year = {2023},
	pages = {44},
}

@article{yu_born_2023,
	title = {Born this way: thin disc, thick disc, and isotropic spheroid formation in {FIRE}-2 {Milky} {Way}-mass galaxy simulations},
	volume = {523},
	issn = {0035-8711},
	shorttitle = {Born this way},
	url = {https://ui.adsabs.harvard.edu/abs/2023MNRAS.523.6220Y},
	doi = {10.1093/mnras/stad1806},
	urldate = {2024-01-30},
	journal = {\mnras},
	author = {Yu, Sijie and Bullock, James S. and Gurvich, Alexander B. and Hafen, Zachary and Stern, Jonathan and Boylan-Kolchin, Michael and Faucher-Giguère, Claude-André and Wetzel, Andrew and Hopkins, Philip F. and Moreno, Jorge},
	month = aug,
	year = {2023},
	pages = {6220--6238},
}

@ARTICLE{fernandez-alvar_metal-poor_2024,
       author = {{Fern{\'a}ndez-Alvar}, Emma and {Kordopatis}, Georges and {Hill}, Vanessa and {Battaglia}, Giuseppina and {Gallart}, Carme and {Gonz{\'a}lez Rivera de la Vernhe}, Isaure and {Thomas}, Guillaume and {Sestito}, Federico and {Ardern-Arentsen}, Anke and {Martin}, Nicolas and {Viswanathan}, Akshara and {Starkenburg}, Else},
        title = "{The metal-poor edge of the Milky Way's ``thin disc''}",
      journal = {\aap},
         year = 2024,
        month = may,
       volume = {685},
          eid = {A151},
        pages = {A151},
          doi = {10.1051/0004-6361/202348918},
archivePrefix = {arXiv},
       eprint = {2402.02943},
 primaryClass = {astro-ph.GA},
       adsurl = {https://ui.adsabs.harvard.edu/abs/2024A&A...685A.151F}
}

@article{belokurov_dawn_2022,
	title = {From dawn till disc: {Milky} {Way}’s turbulent youth revealed by the {APOGEE}+{Gaia} data},
	volume = {514},
	issn = {0035-8711},
	shorttitle = {From dawn till disc},
	url = {https://doi.org/10.1093/mnras/stac1267},
	doi = {10.1093/mnras/stac1267},
	number = {1},
	urldate = {2024-03-15},
	journal = {\mnras},
	author = {Belokurov, Vasily and Kravtsov, Andrey},
	month = jul,
	year = {2022},
	pages = {689--714},
}

@article{pedregosa_scikit-learn_2011,
	title = {Scikit-learn: {Machine} {Learning} in {Python}},
	volume = {12},
	url = {http://jmlr.org/papers/v12/pedregosa11a.html},
	number = {85},
	journal = {Journal of Machine Learning Research},
	author = {Pedregosa, Fabian and Varoquaux, Gaël and Gramfort, Alexandre and Michel, Vincent and Thirion, Bertrand and Grisel, Olivier and Blondel, Mathieu and Prettenhofer, Peter and Weiss, Ron and Dubourg, Vincent and Vanderplas, Jake and Passos, Alexandre and Cournapeau, David and Brucher, Matthieu and Perrot, Matthieu and Duchesnay, Édouard},
	year = {2011},
	pages = {2825--2830},
}

@article{hopkins_new_2015,
	title = {A new class of accurate, mesh-free hydrodynamic simulation methods},
	volume = {450},
	issn = {0035-8711},
	url = {https://doi.org/10.1093/mnras/stv195},
	doi = {10.1093/mnras/stv195},
	number = {1},
	journal = {\mnras},
	author = {Hopkins, Philip F.},
	month = apr,
	year = {2015},
	pages = {53--110},
}

@ARTICLE{gallart_chronology_2024,
       author = {{Gallart}, Carme and {Surot}, Francisco and {Cassisi}, Santi and {Fern{\'a}ndez-Alvar}, Emma and {Mirabal}, David and {Rivero}, Alicia and {Ruiz-Lara}, Tom{\'a}s and {Santos-Torres}, Judith and {Aznar-Menargues}, Guillem and {Battaglia}, Giuseppina and {Queiroz}, Anna B. and {Monelli}, Matteo and {Vasiliev}, Eugene and {Chiappini}, Cristina and {Helmi}, Amina and {Hill}, Vanessa and {Massari}, Davide and {Thomas}, Guillaume F.},
        title = "{Chronology of our Galaxy from Gaia colour-magnitude diagram fitting (ChronoGal). I. The formation and evolution of the thin disc from the Gaia Catalogue of Nearby Stars}",
      journal = {\aap},
         year = 2024,
        month = jul,
       volume = {687},
          eid = {A168},
        pages = {A168},
          doi = {10.1051/0004-6361/202349078},
archivePrefix = {arXiv},
       eprint = {2402.09399},
 primaryClass = {astro-ph.GA},
       adsurl = {https://ui.adsabs.harvard.edu/abs/2024A&A...687A.168G}
}

@article{hopkins_fire-2_2018,
	title = {{FIRE}-2 simulations: physics versus numerics in galaxy formation},
	volume = {480},
	issn = {0035-8711},
	url = {https://doi.org/10.1093/mnras/sty1690},
	doi = {10.1093/mnras/sty1690},
	number = {1},
	journal = {\mnras},
	author = {Hopkins, Philip F and Wetzel, Andrew and Kereš, Dušan and Faucher-Giguère, Claude-André and Quataert, Eliot and Boylan-Kolchin, Michael and Murray, Norman and Hayward, Christopher C and Garrison-Kimmel, Shea and Hummels, Cameron and Feldmann, Robert and Torrey, Paul and Ma, Xiangcheng and Anglés-Alcázar, Daniel and Su, Kung-Yi and Orr, Matthew and Schmitz, Denise and Escala, Ivanna and Sanderson, Robyn and Grudić, Michael Y and Hafen, Zachary and Kim, Ji-Hoon and Fitts, Alex and Bullock, James S and Wheeler, Coral and Chan, T K and Elbert, Oliver D and Narayanan, Desika},
	month = jun,
	year = {2018},
	pages = {800--863},
}

@article{yu_bursty_2021,
	title = {The bursty origin of the {Milky} {Way} thick disc},
	volume = {505},
	copyright = {https://academic.oup.com/journals/pages/open\_access/funder\_policies/chorus/standard\_publication\_model},
	issn = {0035-8711, 1365-2966},
	url = {https://academic.oup.com/mnras/article/505/1/889/6275196},
	doi = {10.1093/mnras/stab1339},
	language = {en},
	number = {1},
	urldate = {2024-04-17},
	journal = {\mnras},
	author = {Yu, Sijie and Bullock, James S and Klein, Courtney and Stern, Jonathan and Wetzel, Andrew and Ma, Xiangcheng and Moreno, Jorge and Hafen, Zachary and Gurvich, Alexander B and Hopkins, Philip F and Kereš, Dušan and Faucher-Giguère, Claude-André and Feldmann, Robert and Quataert, Eliot},
	month = may,
	year = {2021},
	pages = {889--902},
}

@ARTICLE{2015ApJ...808..132H,
       author = {{Hayden}, Michael R. and {Bovy}, Jo and {Holtzman}, Jon A. and {Nidever}, David L. and {Bird}, Jonathan C. and {Weinberg}, David H. and {Andrews}, Brett H. and {Majewski}, Steven R. and {Allende Prieto}, Carlos and {Anders}, Friedrich and {Beers}, Timothy C. and {Bizyaev}, Dmitry and {Chiappini}, Cristina and {Cunha}, Katia and {Frinchaboy}, Peter and {Garc{\'\i}a-Her{\'n}andez}, D.~A. and {Garc{\'\i}a P{\'e}rez}, Ana E. and {Girardi}, L{\'e}o and {Harding}, Paul and {Hearty}, Fred R. and {Johnson}, Jennifer A. and {M{\'e}sz{\'a}ros}, Szabolcs and {Minchev}, Ivan and {O'Connell}, Robert and {Pan}, Kaike and {Robin}, Annie C. and {Schiavon}, Ricardo P. and {Schneider}, Donald P. and {Schultheis}, Mathias and {Shetrone}, Matthew and {Skrutskie}, Michael and {Steinmetz}, Matthias and {Smith}, Verne and {Wilson}, John C. and {Zamora}, Olga and {Zasowski}, Gail},
        title = "{Chemical Cartography with APOGEE: Metallicity Distribution Functions and the Chemical Structure of the Milky Way Disk}",
      journal = {\apj},
         year = 2015,
        month = aug,
       volume = {808},
       number = {2},
          eid = {132},
        pages = {132},
          doi = {10.1088/0004-637X/808/2/132},
archivePrefix = {arXiv},
       eprint = {1503.02110},
 primaryClass = {astro-ph.GA},
       adsurl = {https://ui.adsabs.harvard.edu/abs/2015ApJ...808..132H}
}

@ARTICLE{2023ApJ...954..124I,
       author = {{Imig}, Julie and {Price}, Cathryn and {Holtzman}, Jon A. and {Stone-Martinez}, Alexander and {Majewski}, Steven R. and {Weinberg}, David H. and {Johnson}, Jennifer A. and {Allende Prieto}, Carlos and {Beaton}, Rachael L. and {Beers}, Timothy C. and {Bizyaev}, Dmitry and {Blanton}, Michael R. and {Brownstein}, Joel R. and {Cunha}, Katia and {Fern{\'a}ndez-Trincado}, Jos{\'e} G. and {Feuillet}, Diane K. and {Hasselquist}, Sten and {Hayes}, Christian R. and {J{\"o}nsson}, Henrik and {Lane}, Richard R. and {Lian}, Jianhui and {M{\'e}sz{\'a}ros}, Szabolcs and {Nidever}, David L. and {Robin}, Annie C. and {Shetrone}, Matthew and {Smith}, Verne and {Wilson}, John C.},
        title = "{A Tale of Two Disks: Mapping the Milky Way with the Final Data Release of APOGEE}",
      journal = {\apj},
         year = 2023,
        month = sep,
       volume = {954},
       number = {2},
          eid = {124},
        pages = {124},
          doi = {10.3847/1538-4357/ace9b8},
archivePrefix = {arXiv},
       eprint = {2307.13887},
 primaryClass = {astro-ph.GA},
       adsurl = {https://ui.adsabs.harvard.edu/abs/2023ApJ...954..124I}
}

@ARTICLE{2024A&A...682A...9G,
       author = {{Guiglion}, G. and {Nepal}, S. and {Chiappini}, C. and {Khoperskov}, S. and {Traven}, G. and {Queiroz}, A.~B.~A. and {Steinmetz}, M. and {Valentini}, M. and {Fournier}, Y. and {Vallenari}, A. and {Youakim}, K. and {Bergemann}, M. and {M{\'e}sz{\'a}ros}, S. and {Lucatello}, S. and {Sordo}, R. and {Fabbro}, S. and {Minchev}, I. and {Tautvai{\v{s}}ien{\.{e}}}, G. and {Mikolaitis}, {\v{S}}. and {Montalb{\'a}n}, J.},
        title = "{Beyond Gaia DR3: Tracing the [{\ensuremath{\alpha}}/M] - [M/H] bimodality from the inner to the outer Milky Way disc with Gaia-RVS and convolutional neural networks}",
      journal = {\aap},
         year = 2024,
        month = feb,
       volume = {682},
          eid = {A9},
        pages = {A9},
          doi = {10.1051/0004-6361/202347122},
archivePrefix = {arXiv},
       eprint = {2306.05086},
 primaryClass = {astro-ph.GA},
       adsurl = {https://ui.adsabs.harvard.edu/abs/2024A&A...682A...9G}
}

@ARTICLE{1998A&A...338..161F,
       author = {{Fuhrmann}, Klaus},
        title = "{Nearby stars of the Galactic disk and halo}",
      journal = {\aap},
         year = 1998,
        month = oct,
       volume = {338},
        pages = {161-183},
       adsurl = {https://ui.adsabs.harvard.edu/abs/1998A&A...338..161F}
}

@ARTICLE{2014A&A...562A..71B,
       author = {{Bensby}, T. and {Feltzing}, S. and {Oey}, M.~S.},
        title = "{Exploring the Milky Way stellar disk. A detailed elemental abundance study of 714 F and G dwarf stars in the solar neighbourhood}",
      journal = {\aap},
         year = 2014,
        month = feb,
       volume = {562},
          eid = {A71},
        pages = {A71},
          doi = {10.1051/0004-6361/201322631},
archivePrefix = {arXiv},
       eprint = {1309.2631},
 primaryClass = {astro-ph.GA},
       adsurl = {https://ui.adsabs.harvard.edu/abs/2014A&A...562A..71B}
}

@ARTICLE{1997ApJ...477..765C,
       author = {{Chiappini}, C. and {Matteucci}, F. and {Gratton}, R.},
        title = "{The Chemical Evolution of the Galaxy: The Two-Infall Model}",
      journal = {\apj},
         year = 1997,
        month = mar,
       volume = {477},
       number = {2},
        pages = {765-780},
          doi = {10.1086/303726},
archivePrefix = {arXiv},
       eprint = {astro-ph/9609199},
 primaryClass = {astro-ph},
       adsurl = {https://ui.adsabs.harvard.edu/abs/1997ApJ...477..765C}
}

@ARTICLE{2022A&A...663A.174S,
       author = {{Spitoni}, E. and {Aguirre B{\o}rsen-Koch}, V. and {Verma}, K. and {Stokholm}, A.},
        title = "{Disc dichotomy signature in the vertical distribution of [Mg/Fe] and the delayed gas infall scenario}",
      journal = {\aap},
         year = 2022,
        month = jul,
       volume = {663},
          eid = {A174},
        pages = {A174},
          doi = {10.1051/0004-6361/202142469},
archivePrefix = {arXiv},
       eprint = {2204.07597},
 primaryClass = {astro-ph.GA},
       adsurl = {https://ui.adsabs.harvard.edu/abs/2022A&A...663A.174S}
}

@ARTICLE{2021MNRAS.501.5176K,
       author = {{Khoperskov}, Sergey and {Haywood}, Misha and {Snaith}, Owain and {Di Matteo}, Paola and {Lehnert}, Matthew and {Vasiliev}, Evgenii and {Naroenkov}, Sergey and {Berczik}, Peter},
        title = "{Bimodality of [{\ensuremath{\alpha}} Fe]-[Fe/H] distributions is a natural outcome of dissipative collapse and disc growth in Milky Way-type galaxies}",
      journal = {\mnras},
         year = 2021,
        month = mar,
       volume = {501},
       number = {4},
        pages = {5176-5196},
          doi = {10.1093/mnras/staa3996},
archivePrefix = {arXiv},
       eprint = {2006.10195},
 primaryClass = {astro-ph.GA},
       adsurl = {https://ui.adsabs.harvard.edu/abs/2021MNRAS.501.5176K}
}

@ARTICLE{2002MNRAS.336..785S,
       author = {{Sellwood}, J.~A. and {Binney}, J.~J.},
        title = "{Radial mixing in galactic discs}",
      journal = {\mnras},
         year = 2002,
        month = nov,
       volume = {336},
       number = {3},
        pages = {785-796},
          doi = {10.1046/j.1365-8711.2002.05806.x},
archivePrefix = {arXiv},
       eprint = {astro-ph/0203510},
 primaryClass = {astro-ph},
       adsurl = {https://ui.adsabs.harvard.edu/abs/2002MNRAS.336..785S}
}

@ARTICLE{2009MNRAS.396..203S,
       author = {{Sch{\"o}nrich}, Ralph and {Binney}, James},
        title = "{Chemical evolution with radial mixing}",
      journal = {\mnras},
         year = 2009,
        month = jun,
       volume = {396},
       number = {1},
        pages = {203-222},
          doi = {10.1111/j.1365-2966.2009.14750.x},
archivePrefix = {arXiv},
       eprint = {0809.3006},
 primaryClass = {astro-ph},
       adsurl = {https://ui.adsabs.harvard.edu/abs/2009MNRAS.396..203S}
}

@ARTICLE{2013A&A...558A...9M,
       author = {{Minchev}, I. and {Chiappini}, C. and {Martig}, M.},
        title = "{Chemodynamical evolution of the Milky Way disk. I. The solar vicinity}",
      journal = {\aap},
         year = 2013,
        month = oct,
       volume = {558},
          eid = {A9},
        pages = {A9},
          doi = {10.1051/0004-6361/201220189},
archivePrefix = {arXiv},
       eprint = {1208.1506},
 primaryClass = {astro-ph.GA},
       adsurl = {https://ui.adsabs.harvard.edu/abs/2013A&A...558A...9M}
}

@ARTICLE{2019MNRAS.484.3476C,
       author = {{Clarke}, Adam J. and {Debattista}, Victor P. and {Nidever}, David L. and {Loebman}, Sarah R. and {Simons}, Raymond C. and {Kassin}, Susan and {Du}, Min and {Ness}, Melissa and {Fisher}, Deanne B. and {Quinn}, Thomas R. and {Wadsley}, James and {Freeman}, Ken C. and {Popescu}, Cristina C.},
        title = "{The imprint of clump formation at high redshift - I. A disc {\ensuremath{\alpha}}-abundance dichotomy}",
      journal = {\mnras},
         year = 2019,
        month = apr,
       volume = {484},
       number = {3},
        pages = {3476-3490},
          doi = {10.1093/mnras/stz104},
archivePrefix = {arXiv},
       eprint = {1901.00931},
 primaryClass = {astro-ph.GA},
       adsurl = {https://ui.adsabs.harvard.edu/abs/2019MNRAS.484.3476C}
}

@ARTICLE{1981A&A....98....1M,
       author = {{Mayor}, M. and {Vigroux}, L.},
        title = "{Effect of the infall of matter on the dynamical structure and chemical evolution of a spiral galaxy}",
      journal = {\aap},
         year = 1981,
        month = may,
       volume = {98},
       number = {1},
        pages = {1-8},
       adsurl = {https://ui.adsabs.harvard.edu/abs/1981A&A....98....1M}
}

@ARTICLE{2019MNRAS.489.4233M,
       author = {{Marinacci}, Federico and {Sales}, Laura V. and {Vogelsberger}, Mark and {Torrey}, Paul and {Springel}, Volker},
        title = "{Simulating the interstellar medium and stellar feedback on a moving mesh: implementation and isolated galaxies}",
      journal = {\mnras},
         year = 2019,
        month = nov,
       volume = {489},
       number = {3},
        pages = {4233-4260},
          doi = {10.1093/mnras/stz2391},
archivePrefix = {arXiv},
       eprint = {1905.08806},
 primaryClass = {astro-ph.GA},
       adsurl = {https://ui.adsabs.harvard.edu/abs/2019MNRAS.489.4233M}
}

@article{mccluskey_disc_2024,
	title = {Disc settling and dynamical heating: histories of {Milky} {Way}-mass stellar discs across cosmic time in the {FIRE} simulations},
	volume = {527},
	issn = {0035-8711},
	shorttitle = {Disc settling and dynamical heating},
	url = {https://ui.adsabs.harvard.edu/abs/2024MNRAS.527.6926M},
	doi = {10.1093/mnras/stad3547},
	urldate = {2024-04-19},
	journal = {\mnras},
	author = {McCluskey, Fiona and Wetzel, Andrew and Loebman, Sarah R. and Moreno, Jorge and Faucher-Giguère, Claude-André and Hopkins, Philip F.},
	month = jan,
	year = {2024},
	pages = {6926--6949},
}

@ARTICLE{2020JCAP...05..033K,
       author = {{Karukes}, E.~V. and {Benito}, M. and {Iocco}, F. and {Trotta}, R. and {Geringer-Sameth}, A.},
        title = "{A robust estimate of the Milky Way mass from rotation curve data}",
      journal = {\jcap},
         year = 2020,
        month = may,
       volume = {2020},
       number = {5},
          eid = {033},
        pages = {033},
          doi = {10.1088/1475-7516/2020/05/033},
archivePrefix = {arXiv},
       eprint = {1912.04296},
 primaryClass = {astro-ph.GA},
       adsurl = {https://ui.adsabs.harvard.edu/abs/2020JCAP...05..033K}
}

@ARTICLE{2018MNRAS.480..800H,
       author = {{Hopkins}, Philip F. and {Wetzel}, Andrew and {Kere{\v{s}}}, Du{\v{s}}an and {Faucher-Gigu{\`e}re}, Claude-Andr{\'e} and {Quataert}, Eliot and {Boylan-Kolchin}, Michael and {Murray}, Norman and {Hayward}, Christopher C. and {Garrison-Kimmel}, Shea and {Hummels}, Cameron and {Feldmann}, Robert and {Torrey}, Paul and {Ma}, Xiangcheng and {Angl{\'e}s-Alc{\'a}zar}, Daniel and {Su}, Kung-Yi and {Orr}, Matthew and {Schmitz}, Denise and {Escala}, Ivanna and {Sanderson}, Robyn and {Grudi{\'c}}, Michael Y. and {Hafen}, Zachary and {Kim}, Ji-Hoon and {Fitts}, Alex and {Bullock}, James S. and {Wheeler}, Coral and {Chan}, T.~K. and {Elbert}, Oliver D. and {Narayanan}, Desika},
        title = "{FIRE-2 simulations: physics versus numerics in galaxy formation}",
      journal = {\mnras},
         year = 2018,
        month = oct,
       volume = {480},
       number = {1},
        pages = {800-863},
          doi = {10.1093/mnras/sty1690},
archivePrefix = {arXiv},
       eprint = {1702.06148},
 primaryClass = {astro-ph.GA},
       adsurl = {https://ui.adsabs.harvard.edu/abs/2018MNRAS.480..800H}
}

@ARTICLE{2019MNRAS.487.1380G,
       author = {{Garrison-Kimmel}, Shea and {Hopkins}, Philip F. and {Wetzel}, Andrew and {Bullock}, James S. and {Boylan-Kolchin}, Michael and {Kere{\v{s}}}, Du{\v{s}}an and {Faucher-Gigu{\`e}re}, Claude-Andr{\'e} and {El-Badry}, Kareem and {Lamberts}, Astrid and {Quataert}, Eliot and {Sanderson}, Robyn},
        title = "{The Local Group on FIRE: dwarf galaxy populations across a suite of hydrodynamic simulations}",
      journal = {\mnras},
         year = 2019,
        month = jul,
       volume = {487},
       number = {1},
        pages = {1380-1399},
          doi = {10.1093/mnras/stz1317},
archivePrefix = {arXiv},
       eprint = {1806.04143},
 primaryClass = {astro-ph.GA},
       adsurl = {https://ui.adsabs.harvard.edu/abs/2019MNRAS.487.1380G}
}

@ARTICLE{2025ApJ...978...37A,
       author = {{Ansar}, Sioree and {Pearson}, Sarah and {Sanderson}, Robyn E. and {Arora}, Arpit and {Hopkins}, Philip F. and {Wetzel}, Andrew and {Cunningham}, Emily C. and {Quinn}, Jamie},
        title = "{Bar Formation and Destruction in the FIRE-2 Simulations}",
      journal = {\apj},
         year = 2025,
        month = jan,
       volume = {978},
       number = {1},
          eid = {37},
        pages = {37},
          doi = {10.3847/1538-4357/ad8b45},
archivePrefix = {arXiv},
       eprint = {2309.16811},
 primaryClass = {astro-ph.GA},
       adsurl = {https://ui.adsabs.harvard.edu/abs/2025ApJ...978...37A}
}

@ARTICLE{2019MNRAS.482.3089N,
       author = {{Nuza}, Sebasti{\'a}n E. and {Scannapieco}, Cecilia and {Chiappini}, Cristina and {Junqueira}, Thiago C. and {Minchev}, Ivan and {Martig}, Marie},
        title = "{Gas accretion in Milky Way-like galaxies: temporal and radial dependencies}",
      journal = {\mnras},
         year = 2019,
        month = jan,
       volume = {482},
       number = {3},
        pages = {3089-3108},
          doi = {10.1093/mnras/sty2882},
archivePrefix = {arXiv},
       eprint = {1805.06428},
 primaryClass = {astro-ph.GA},
       adsurl = {https://ui.adsabs.harvard.edu/abs/2019MNRAS.482.3089N}
}

@ARTICLE{2020A&A...640A..81N,
       author = {{Nissen}, P.~E. and {Christensen-Dalsgaard}, J. and {Mosumgaard}, J.~R. and {Silva Aguirre}, V. and {Spitoni}, E. and {Verma}, K.},
        title = "{High-precision abundances of elements in solar-type stars. Evidence of two distinct sequences in abundance-age relations}",
      journal = {\aap},
         year = 2020,
        month = aug,
       volume = {640},
          eid = {A81},
        pages = {A81},
          doi = {10.1051/0004-6361/202038300},
archivePrefix = {arXiv},
       eprint = {2006.06013},
 primaryClass = {astro-ph.SR},
       adsurl = {https://ui.adsabs.harvard.edu/abs/2020A&A...640A..81N}
}

@ARTICLE{2015ApJ...800...14M,
       author = {{Miller}, Matthew J. and {Bregman}, Joel N.},
        title = "{Constraining the Milky Way's Hot Gas Halo with O VII and O VIII Emission Lines}",
      journal = {\apj},
         year = 2015,
        month = feb,
       volume = {800},
       number = {1},
          eid = {14},
        pages = {14},
          doi = {10.1088/0004-637X/800/1/14},
archivePrefix = {arXiv},
       eprint = {1412.3116},
 primaryClass = {astro-ph.GA},
       adsurl = {https://ui.adsabs.harvard.edu/abs/2015ApJ...800...14M}
}

@ARTICLE{2016ARA&A..54..529B,
       author = {{Bland-Hawthorn}, Joss and {Gerhard}, Ortwin},
        title = "{The Galaxy in Context: Structural, Kinematic, and Integrated Properties}",
      journal = {\araa},
         year = 2016,
        month = sep,
       volume = {54},
        pages = {529-596},
          doi = {10.1146/annurev-astro-081915-023441},
archivePrefix = {arXiv},
       eprint = {1602.07702},
 primaryClass = {astro-ph.GA},
       adsurl = {https://ui.adsabs.harvard.edu/abs/2016ARA&A..54..529B}
}

@ARTICLE{2011MNRAS.414.2446M,
       author = {{McMillan}, Paul J.},
        title = "{Mass models of the Milky Way}",
      journal = {\mnras},
         year = 2011,
        month = jul,
       volume = {414},
       number = {3},
        pages = {2446-2457},
          doi = {10.1111/j.1365-2966.2011.18564.x},
archivePrefix = {arXiv},
       eprint = {1102.4340},
 primaryClass = {astro-ph.GA},
       adsurl = {https://ui.adsabs.harvard.edu/abs/2011MNRAS.414.2446M}
}

@ARTICLE{2024NatAs...8.1302L,
       author = {{Lian}, Jianhui and {Zasowski}, Gail and {Chen}, Bingqiu and {Imig}, Julie and {Wang}, Tao and {Boardman}, Nicholas and {Liu}, Xiaowei},
        title = "{The broken-exponential radial structure and larger size of the Milky Way galaxy}",
      journal = {Nature Astronomy},
         year = 2024,
        month = oct,
       volume = {8},
       number = {10},
        pages = {1302-1309},
          doi = {10.1038/s41550-024-02315-7},
archivePrefix = {arXiv},
       eprint = {2406.05604},
 primaryClass = {astro-ph.GA},
       adsurl = {https://ui.adsabs.harvard.edu/abs/2024NatAs...8.1302L}
}

@ARTICLE{2021MNRAS.503.5826A,
       author = {{Agertz}, Oscar and {Renaud}, Florent and {Feltzing}, Sofia and {Read}, Justin I. and {Ryde}, Nils and {Andersson}, Eric P. and {Rey}, Martin P. and {Bensby}, Thomas and {Feuillet}, Diane K.},
        title = "{VINTERGATAN - I. The origins of chemically, kinematically, and structurally distinct discs in a simulated Milky Way-mass galaxy}",
      journal = {\mnras},
         year = 2021,
        month = jun,
       volume = {503},
       number = {4},
        pages = {5826-5845},
          doi = {10.1093/mnras/stab322},
archivePrefix = {arXiv},
       eprint = {2006.06008},
 primaryClass = {astro-ph.GA},
       adsurl = {https://ui.adsabs.harvard.edu/abs/2021MNRAS.503.5826A}
}

@ARTICLE{1976MNRAS.176...31L,
       author = {{Larson}, R.~B.},
        title = "{Models for the formation of disc galaxies.}",
      journal = {\mnras},
         year = 1976,
        month = jul,
       volume = {176},
        pages = {31-52},
          doi = {10.1093/mnras/176.1.31},
       adsurl = {https://ui.adsabs.harvard.edu/abs/1976MNRAS.176...31L}
}

@ARTICLE{2022MNRAS.514..689B,
       author = {{Belokurov}, Vasily and {Kravtsov}, Andrey},
        title = "{From dawn till disc: Milky Way's turbulent youth revealed by the APOGEE+Gaia data}",
      journal = {\mnras},
         year = 2022,
        month = jul,
       volume = {514},
       number = {1},
        pages = {689-714},
          doi = {10.1093/mnras/stac1267},
archivePrefix = {arXiv},
       eprint = {2203.04980},
 primaryClass = {astro-ph.GA},
       adsurl = {https://ui.adsabs.harvard.edu/abs/2022MNRAS.514..689B}
}

@ARTICLE{bay_gaia,
       author = {{P{\~o}der}, Sven and {Benito}, Mar{\'\i}a and {Pata}, Joosep and {Kipper}, Rain and {Ramler}, Heleri and {H{\"u}tsi}, Gert and {Kolka}, Indrek and {Thomas}, Guillaume F.},
        title = "{A Bayesian estimation of the Milky Way's circular velocity curve using Gaia DR3}",
      journal = {\aap},
         year = 2023,
        month = aug,
       volume = {676},
          eid = {A134},
        pages = {A134},
          doi = {10.1051/0004-6361/202346474},
archivePrefix = {arXiv},
       eprint = {2309.02895},
 primaryClass = {astro-ph.GA},
       adsurl = {https://ui.adsabs.harvard.edu/abs/2023A&A...676A.134P}
}

@ARTICLE{gspphot,
       author = {{Andrae}, R. and {Fouesneau}, M. and {Sordo}, R. and {Bailer-Jones}, C.~A.~L. and {Dharmawardena}, T.~E. and {Rybizki}, J. and {De Angeli}, F. and {Lindstr{\o}m}, H.~E.~P. and {Marshall}, D.~J. and {Drimmel}, R. and {Korn}, A.~J. and {Soubiran}, C. and {Brouillet}, N. and {Casamiquela}, L. and {Rix}, H. -W. and {Abreu Aramburu}, A. and {{\'A}lvarez}, M.~A. and {Bakker}, J. and {Bellas-Velidis}, I. and {Bijaoui}, A. and {Brugaletta}, E. and {Burlacu}, A. and {Carballo}, R. and {Chaoul}, L. and {Chiavassa}, A. and {Contursi}, G. and {Cooper}, W.~J. and {Creevey}, O.~L. and {Dafonte}, C. and {Dapergolas}, A. and {de Laverny}, P. and {Delchambre}, L. and {Demouchy}, C. and {Edvardsson}, B. and {Fr{\'e}mat}, Y. and {Garabato}, D. and {Garc{\'\i}a-Lario}, P. and {Garc{\'\i}a-Torres}, M. and {Gavel}, A. and {Gomez}, A. and {Gonz{\'a}lez-Santamar{\'\i}a}, I. and {Hatzidimitriou}, D. and {Heiter}, U. and {Jean-Antoine Piccolo}, A. and {Kontizas}, M. and {Kordopatis}, G. and {Lanzafame}, A.~C. and {Lebreton}, Y. and {Licata}, E.~L. and {Livanou}, E. and {Lobel}, A. and {Lorca}, A. and {Magdaleno Romeo}, A. and {Manteiga}, M. and {Marocco}, F. and {Mary}, N. and {Nicolas}, C. and {Ordenovic}, C. and {Pailler}, F. and {Palicio}, P.~A. and {Pallas-Quintela}, L. and {Panem}, C. and {Pichon}, B. and {Poggio}, E. and {Recio-Blanco}, A. and {Riclet}, F. and {Robin}, C. and {Santove{\~n}a}, R. and {Sarro}, L.~M. and {Schultheis}, M.~S. and {Segol}, M. and {Silvelo}, A. and {Slezak}, I. and {Smart}, R.~L. and {S{\"u}veges}, M. and {Th{\'e}venin}, F. and {Torralba Elipe}, G. and {Ulla}, A. and {Utrilla}, E. and {Vallenari}, A. and {van Dillen}, E. and {Zhao}, H. and {Zorec}, J.},
        title = "{Gaia Data Release 3. Analysis of the Gaia BP/RP spectra using the General Stellar Parameterizer from Photometry}",
      journal = {\aap},
         year = 2023,
        month = jun,
       volume = {674},
          eid = {A27},
        pages = {A27},
          doi = {10.1051/0004-6361/202243462},
archivePrefix = {arXiv},
       eprint = {2206.06138},
 primaryClass = {astro-ph.SR},
       adsurl = {https://ui.adsabs.harvard.edu/abs/2023A&A...674A..27A}
}

@ARTICLE{2021ApJS..254....2P,
       author = {{Park}, Minjung J. and {Yi}, Sukyoung K. and {Peirani}, Sebastien and {Pichon}, Christophe and {Dubois}, Yohan and {Choi}, Hoseung and {Devriendt}, Julien and {Kaviraj}, Sugata and {Kimm}, Taysun and {Kraljic}, Katarina and {Volonteri}, Marta},
        title = "{Exploring the Origin of Thick Disks Using the NewHorizon and Galactica Simulations}",
      journal = {\apjs},
         year = 2021,
        month = may,
       volume = {254},
       number = {1},
          eid = {2},
        pages = {2},
          doi = {10.3847/1538-4365/abe937},
archivePrefix = {arXiv},
       eprint = {2009.12373},
 primaryClass = {astro-ph.GA},
       adsurl = {https://ui.adsabs.harvard.edu/abs/2021ApJS..254....2P}
}

@ARTICLE{2021MNRAS.503.2814C,
       author = {{Ciuc{\u{a}}}, Ioana and {Kawata}, Daisuke and {Miglio}, Andrea and {Davies}, Guy R. and {Grand}, Robert J.~J.},
        title = "{Unveiling the distinct formation pathways of the inner and outer discs of the Milky Way with Bayesian Machine Learning}",
      journal = {\mnras},
         year = 2021,
        month = may,
       volume = {503},
       number = {2},
        pages = {2814-2824},
          doi = {10.1093/mnras/stab639},
archivePrefix = {arXiv},
       eprint = {2003.03316},
 primaryClass = {astro-ph.GA},
       adsurl = {https://ui.adsabs.harvard.edu/abs/2021MNRAS.503.2814C}
}

@ARTICLE{2004ApJ...612..894B,
       author = {{Brook}, Chris B. and {Kawata}, Daisuke and {Gibson}, Brad K. and {Freeman}, Ken C.},
        title = "{The Emergence of the Thick Disk in a Cold Dark Matter Universe}",
      journal = {\apj},
         year = 2004,
        month = sep,
       volume = {612},
       number = {2},
        pages = {894-899},
          doi = {10.1086/422709},
archivePrefix = {arXiv},
       eprint = {astro-ph/0405306},
 primaryClass = {astro-ph},
       adsurl = {https://ui.adsabs.harvard.edu/abs/2004ApJ...612..894B}
}

@ARTICLE{2024A&A...691A..61P,
       author = {{Pinna}, Francesca and {Grand}, Robert J.~J. and {Martig}, Marie and {Fragkoudi}, Francesca},
        title = "{Recovering chemical bimodalities in observed edge-on stellar disks: Insights from AURIGA simulations}",
      journal = {\aap},
         year = 2024,
        month = nov,
       volume = {691},
          eid = {A61},
        pages = {A61},
          doi = {10.1051/0004-6361/202450843},
archivePrefix = {arXiv},
       eprint = {2409.07533},
 primaryClass = {astro-ph.GA},
       adsurl = {https://ui.adsabs.harvard.edu/abs/2024A&A...691A..61P}
}

@ARTICLE{2023ApJ...956L..14K,
       author = {{Kobayashi}, Chiaki and {Bhattacharya}, Souradeep and {Arnaboldi}, Magda and {Gerhard}, Ortwin},
        title = "{On the {\ensuremath{\alpha}}/Fe Bimodality of the M31 Disks}",
      journal = {\apjl},
         year = 2023,
        month = oct,
       volume = {956},
       number = {1},
          eid = {L14},
        pages = {L14},
          doi = {10.3847/2041-8213/acf7c7},
archivePrefix = {arXiv},
       eprint = {2309.01707},
 primaryClass = {astro-ph.GA},
       adsurl = {https://ui.adsabs.harvard.edu/abs/2023ApJ...956L..14K}
}

@Article{matplotlib,
  Author    = {Hunter, J. D.},
  Title     = {Matplotlib: A 2D graphics environment},
  Journal   = {Computing in Science \& Engineering},
  Volume    = {9},
  Number    = {3},
  Pages     = {90--95},
  publisher = {IEEE COMPUTER SOC},
  doi       = {10.1109/MCSE.2007.55},
  year      = 2007
}

@Article{numpy,
 title         = {Array programming with {NumPy}},
 author        = {Charles R. Harris and K. Jarrod Millman and St{\'{e}}fan J.
                 van der Walt and Ralf Gommers and Pauli Virtanen and David
                 Cournapeau and Eric Wieser and Julian Taylor and Sebastian
                 Berg and Nathaniel J. Smith and Robert Kern and Matti Picus
                 and Stephan Hoyer and Marten H. van Kerkwijk and Matthew
                 Brett and Allan Haldane and Jaime Fern{\'{a}}ndez del
                 R{\'{i}}o and Mark Wiebe and Pearu Peterson and Pierre
                 G{\'{e}}rard-Marchant and Kevin Sheppard and Tyler Reddy and
                 Warren Weckesser and Hameer Abbasi and Christoph Gohlke and
                 Travis E. Oliphant},
 year          = {2020},
 month         = sep,
 journal       = {\nat},
 volume        = {585},
 number        = {7825},
 pages         = {357--362},
 doi           = {10.1038/s41586-020-2649-2},
 publisher     = {Springer Science and Business Media {LLC}},
 url           = {https://doi.org/10.1038/s41586-020-2649-2}
}

@conference{jupyter,
Title = {Jupyter Notebooks -- a publishing format for reproducible computational workflows},
Author = {Thomas Kluyver and Benjamin Ragan-Kelley and Fernando P{\'e}rez and Brian Granger and Matthias Bussonnier and Jonathan Frederic and Kyle Kelley and Jessica Hamrick and Jason Grout and Sylvain Corlay and Paul Ivanov and Dami{\'a}n Avila and Safia Abdalla and Carol Willing},
Booktitle = {Positioning and Power in Academic Publishing: Players, Agents and Agendas},
Editor = {F. Loizides and B. Schmidt},
Organization = {IOS Press},
Pages = {87 - 90},
Year = {2016}
}

@ARTICLE{2020SciPy-NMeth,
  author  = {Virtanen, Pauli and Gommers, Ralf and Oliphant, Travis E. and
            Haberland, Matt and Reddy, Tyler and Cournapeau, David and
            Burovski, Evgeni and Peterson, Pearu and Weckesser, Warren and
            Bright, Jonathan and {van der Walt}, St{\'e}fan J. and
            Brett, Matthew and Wilson, Joshua and Millman, K. Jarrod and
            Mayorov, Nikolay and Nelson, Andrew R. J. and Jones, Eric and
            Kern, Robert and Larson, Eric and Carey, C J and
            Polat, {\.I}lhan and Feng, Yu and Moore, Eric W. and
            {VanderPlas}, Jake and Laxalde, Denis and Perktold, Josef and
            Cimrman, Robert and Henriksen, Ian and Quintero, E. A. and
            Harris, Charles R. and Archibald, Anne M. and
            Ribeiro, Ant{\^o}nio H. and Pedregosa, Fabian and
            {van Mulbregt}, Paul and {SciPy 1.0 Contributors}},
  title   = {{{SciPy} 1.0: Fundamental Algorithms for Scientific
            Computing in Python}},
  journal = {Nature Methods},
  year    = {2020},
  volume  = {17},
  pages   = {261--272},
  adsurl  = {https://rdcu.be/b08Wh},
  doi     = {10.1038/s41592-019-0686-2},
}

@ARTICLE{2016ApJ...827L..23W,
       author = {{Wetzel}, Andrew R. and {Hopkins}, Philip F. and {Kim}, Ji-hoon and {Faucher-Gigu{\`e}re}, Claude-Andr{\'e} and {Kere{\v{s}}}, Du{\v{s}}an and {Quataert}, Eliot},
        title = "{Reconciling Dwarf Galaxies with {\ensuremath{\Lambda}}CDM Cosmology: Simulating a Realistic Population of Satellites around a Milky Way-mass Galaxy}",
      journal = {\apjl},
         year = 2016,
        month = aug,
       volume = {827},
       number = {2},
          eid = {L23},
        pages = {L23},
          doi = {10.3847/2041-8205/827/2/L23},
archivePrefix = {arXiv},
       eprint = {1602.05957},
 primaryClass = {astro-ph.GA},
       adsurl = {https://ui.adsabs.harvard.edu/abs/2016ApJ...827L..23W}
}

@INPROCEEDINGS{1978IAUS...79..241J,
       author = {{Joeveer}, M. and {Einasto}, J.},
        title = "{Has the Universe the Cell Structure?}",
    booktitle = {Large Scale Structures in the Universe},
         year = 1978,
       editor = {{Longair}, M.~S. and {Einasto}, J.},
       series = {IAU Symposium},
       volume = {79},
        month = jan,
        pages = {241},
       adsurl = {https://ui.adsabs.harvard.edu/abs/1978IAUS...79..241J}
}

@ARTICLE{1980Natur.283...47E,
       author = {{Einasto}, J. and {Joeveer}, M. and {Saar}, E.},
        title = "{Superclusters and galaxy formation}",
      journal = {\nat},
         year = 1980,
        month = jan,
       volume = {283},
       number = {5742},
        pages = {47-48},
          doi = {10.1038/283047a0},
       adsurl = {https://ui.adsabs.harvard.edu/abs/1980Natur.283...47E}
}

@ARTICLE{2014A&A...567A..68D,
       author = {{Di Teodoro}, E.~M. and {Fraternali}, F.},
        title = "{Gas accretion from minor mergers in local spiral galaxies}",
      journal = {\aap},
         year = 2014,
        month = jul,
       volume = {567},
          eid = {A68},
        pages = {A68},
          doi = {10.1051/0004-6361/201423596},
archivePrefix = {arXiv},
       eprint = {1406.0856},
 primaryClass = {astro-ph.GA},
       adsurl = {https://ui.adsabs.harvard.edu/abs/2014A&A...567A..68D}
}

@ARTICLE{2022Natur.603..599X,
       author = {{Xiang}, Maosheng and {Rix}, Hans-Walter},
        title = "{A time-resolved picture of our Milky Way's early formation history}",
      journal = {\nat},
         year = 2022,
        month = mar,
       volume = {603},
       number = {7902},
        pages = {599-603},
          doi = {10.1038/s41586-022-04496-5},
archivePrefix = {arXiv},
       eprint = {2203.12110},
 primaryClass = {astro-ph.GA},
       adsurl = {https://ui.adsabs.harvard.edu/abs/2022Natur.603..599X}
}

@ARTICLE{2025A&A...701A.270K,
       author = {{Khanna}, Shourya and {Yu}, Jie and {Drimmel}, Ronald and {Poggio}, Eloisa and {Cantat-Gaudin}, Tristan and {Castro-Ginard}, Alfred and {Kurbatov}, Evgeny and {Belokurov}, Vasily and {Brown}, Anthony and {Fouesneau}, Morgan and {Casey}, Andrew and {Rix}, Hans-Walter},
        title = "{GaiaUnlimited: The old stellar disc of the Milky Way as traced by the red clump}",
      journal = {\aap},
         year = 2025,
        month = sep,
       volume = {701},
          eid = {A270},
        pages = {A270},
          doi = {10.1051/0004-6361/202452798},
archivePrefix = {arXiv},
       eprint = {2410.22036},
 primaryClass = {astro-ph.GA},
       adsurl = {https://ui.adsabs.harvard.edu/abs/2025A&A...701A.270K}
}

@ARTICLE{2025A&A...704A.258F,
       author = {{Fern{\'a}ndez-Alvar}, Emma and {Ruiz-Lara}, Tom{\'a}s and {Gallart}, Carme and {Cassisi}, Santi and {Surot}, Francisco and {Gonz{\'a}lez-Koda}, Yllari K. and {Callingham}, Thomas M. and {Queiroz}, Anna B. and {Battaglia}, Giuseppina and {Thomas}, Guillaume and {Chiappini}, Cristina and {Hill}, Vanessa and {Dodd}, Emma and {Helmi}, Amina and {Aznar-Menargues}, Guillem and {de la Cueva}, Alejandro and {Mirabal}, David and {Quintana-Ansaldo}, M{\'o}nica and {Rivero}, Alicia},
        title = "{Chronology of our Galaxy from Gaia colour─magnitude diagram fitting (ChronoGal): II. Unveiling the formation and evolution of the kinematically selected thick and thin discs}",
      journal = {\aap},
         year = 2025,
        month = dec,
       volume = {704},
          eid = {A258},
        pages = {A258},
          doi = {10.1051/0004-6361/202553814},
archivePrefix = {arXiv},
       eprint = {2503.19536},
 primaryClass = {astro-ph.GA},
       adsurl = {https://ui.adsabs.harvard.edu/abs/2025A&A...704A.258F}
}

@ARTICLE{2020A&A...638A..76Q,
       author = {{Queiroz}, A.~B.~A. and {Anders}, F. and {Chiappini}, C. and {Khalatyan}, A. and {Santiago}, B.~X. and {Steinmetz}, M. and {Valentini}, M. and {Miglio}, A. and {Bossini}, D. and {Barbuy}, B. and {Minchev}, I. and {Minniti}, D. and {Garc{\'\i}a Hern{\'a}ndez}, D.~A. and {Schultheis}, M. and {Beaton}, R.~L. and {Beers}, T.~C. and {Bizyaev}, D. and {Brownstein}, J.~R. and {Cunha}, K. and {Fern{\'a}ndez-Trincado}, J.~G. and {Frinchaboy}, P.~M. and {Lane}, R.~R. and {Majewski}, S.~R. and {Nataf}, D. and {Nitschelm}, C. and {Pan}, K. and {Roman-Lopes}, A. and {Sobeck}, J.~S. and {Stringfellow}, G. and {Zamora}, O.},
        title = "{From the bulge to the outer disc: StarHorse stellar parameters, distances, and extinctions for stars in APOGEE DR16 and other spectroscopic surveys}",
      journal = {\aap},
         year = 2020,
        month = jun,
       volume = {638},
          eid = {A76},
        pages = {A76},
          doi = {10.1051/0004-6361/201937364},
archivePrefix = {arXiv},
       eprint = {1912.09778},
 primaryClass = {astro-ph.GA},
       adsurl = {https://ui.adsabs.harvard.edu/abs/2020A&A...638A..76Q}
}

@ARTICLE{2013A&A...560A.109H,
       author = {{Haywood}, Misha and {Di Matteo}, Paola and {Lehnert}, Matthew D. and {Katz}, David and {G{\'o}mez}, Ana},
        title = "{The age structure of stellar populations in the solar vicinity. Clues of a two-phase formation history of the Milky Way disk}",
      journal = {\aap},
         year = 2013,
        month = dec,
       volume = {560},
          eid = {A109},
        pages = {A109},
          doi = {10.1051/0004-6361/201321397},
archivePrefix = {arXiv},
       eprint = {1305.4663},
 primaryClass = {astro-ph.GA},
       adsurl = {https://ui.adsabs.harvard.edu/abs/2013A&A...560A.109H}
}

@ARTICLE{2003A&A...410..527B,
       author = {{Bensby}, T. and {Feltzing}, S. and {Lundstr{\"o}m}, I.},
        title = "{Elemental abundance trends in the Galactic thin and thick disks as traced by nearby F and G dwarf stars}",
      journal = {\aap},
         year = 2003,
        month = nov,
       volume = {410},
        pages = {527-551},
          doi = {10.1051/0004-6361:20031213},
       adsurl = {https://ui.adsabs.harvard.edu/abs/2003A&A...410..527B}
}

@ARTICLE{2022MNRAS.509.4149T,
       author = {{Trapp}, Cameron W. and {Kere{\v{s}}}, Du{\v{s}}an and {Chan}, Tsang Keung and {Escala}, Ivanna and {Hummels}, Cameron and {Hopkins}, Philip F. and {Faucher-Gigu{\`e}re}, Claude-Andr{\'e} and {Murray}, Norman and {Quataert}, Eliot and {Wetzel}, Andrew},
        title = "{Gas infall and radial transport in cosmological simulations of milky way-mass discs}",
      journal = {\mnras},
         year = 2022,
        month = jan,
       volume = {509},
       number = {3},
        pages = {4149-4170},
          doi = {10.1093/mnras/stab3251},
archivePrefix = {arXiv},
       eprint = {2105.11472},
 primaryClass = {astro-ph.GA},
       adsurl = {https://ui.adsabs.harvard.edu/abs/2022MNRAS.509.4149T}
}

@ARTICLE{2022MNRAS.517..832I,
       author = {{Iza}, Federico G. and {Scannapieco}, Cecilia and {Nuza}, Sebasti{\'a}n E. and {Grand}, Robert J.~J. and {G{\'o}mez}, Facundo A. and {Springel}, Volker and {Pakmor}, R{\"u}diger and {Marinacci}, Federico},
        title = "{Cosmological gas accretion history onto the stellar discs of Milky Way-like galaxies in the Auriga simulations - (I) Temporal dependency}",
      journal = {\mnras},
         year = 2022,
        month = nov,
       volume = {517},
       number = {1},
        pages = {832-852},
          doi = {10.1093/mnras/stac2709},
archivePrefix = {arXiv},
       eprint = {2210.03157},
 primaryClass = {astro-ph.GA},
       adsurl = {https://ui.adsabs.harvard.edu/abs/2022MNRAS.517..832I}
}

@ARTICLE{2024ApJ...962...84S,
       author = {{Semenov}, Vadim A. and {Conroy}, Charlie and {Chandra}, Vedant and {Hernquist}, Lars and {Nelson}, Dylan},
        title = "{Formation of Galactic Disks. I. Why Did the Milky Way's Disk Form Unusually Early?}",
      journal = {\apj},
         year = 2024,
        month = feb,
       volume = {962},
       number = {1},
          eid = {84},
        pages = {84},
          doi = {10.3847/1538-4357/ad150a},
archivePrefix = {arXiv},
       eprint = {2306.09398},
 primaryClass = {astro-ph.GA},
       adsurl = {https://ui.adsabs.harvard.edu/abs/2024ApJ...962...84S}
}

@ARTICLE{2026MNRAS.545f1551O,
       author = {{Orkney}, Matthew D.~A. and {Laporte}, Chervin F.~P. and {Grand}, Robert J.~J. and {Springel}, Volker},
        title = "{The Milky Way in context: the formation of galactic discs and chemical sequences from a cosmological perspective}",
      journal = {\mnras},
         year = 2026,
        month = jan,
       volume = {545},
       number = {1},
          eid = {staf1551},
        pages = {staf1551},
          doi = {10.1093/mnras/staf1551},
archivePrefix = {arXiv},
       eprint = {2506.07038},
 primaryClass = {astro-ph.GA},
       adsurl = {https://ui.adsabs.harvard.edu/abs/2026MNRAS.545f1551O}
}

@ARTICLE{2023MNRAS.520.1672P,
       author = {{Parul}, Hanna and {Bailin}, Jeremy and {Wetzel}, Andrew and {Gurvich}, Alexander B. and {Faucher-Gigu{\`e}re}, Claude-Andr{\'e} and {Hafen}, Zachary and {Stern}, Jonathan and {Snaith}, Owain},
        title = "{The imprint of bursty star formation on alpha-element abundance patterns in Milky Way-like galaxies}",
      journal = {\mnras},
         year = 2023,
        month = apr,
       volume = {520},
       number = {2},
        pages = {1672-1686},
          doi = {10.1093/mnras/stad206},
archivePrefix = {arXiv},
       eprint = {2301.07692},
 primaryClass = {astro-ph.GA},
       adsurl = {https://ui.adsabs.harvard.edu/abs/2023MNRAS.520.1672P}
}

@ARTICLE{2025MNRAS.537.1571P,
       author = {{Parul}, Hanna and {Bailin}, Jeremy and {Loebman}, Sarah R. and {Wetzel}, Andrew and {Barry}, Megan and {Bhattarai}, Binod},
        title = "{Effect of gas accretion on {\ensuremath{\alpha}}-element bimodality in Milky Way-mass galaxies in the FIRE-2 simulations}",
      journal = {\mnras},
         year = 2025,
        month = feb,
       volume = {537},
       number = {2},
        pages = {1571-1585},
          doi = {10.1093/mnras/staf137},
archivePrefix = {arXiv},
       eprint = {2501.12342},
 primaryClass = {astro-ph.GA},
       adsurl = {https://ui.adsabs.harvard.edu/abs/2025MNRAS.537.1571P}
}

@ARTICLE{2025arXiv250611840M,
       author = {{McCluskey}, Fiona and {Wetzel}, Andrew and {Loebman}, Sarah and {Moreno}, Jorge},
        title = "{Stellar Velocity Dispersion versus Age: Consistency across Observations and Simulations, with the Milky Way as an Outlier}",
      journal = {arXiv e-prints},
         year = 2025,
        month = jun,
          eid = {arXiv:2506.11840},
        pages = {arXiv:2506.11840},
          doi = {10.48550/arXiv.2506.11840},
archivePrefix = {arXiv},
       eprint = {2506.11840},
 primaryClass = {astro-ph.GA},
       adsurl = {https://ui.adsabs.harvard.edu/abs/2025arXiv250611840M}
}

@ARTICLE{2023MNRAS.524.4091B,
       author = {{Barbani}, Filippo and {Pascale}, Raffaele and {Marinacci}, Federico and {Sales}, Laura V. and {Vogelsberger}, Mark and {Torrey}, Paul and {Li}, Hui},
        title = "{Galactic coronae in Milky Way-like galaxies: the role of stellar feedback in gas accretion}",
      journal = {\mnras},
         year = 2023,
        month = sep,
       volume = {524},
       number = {3},
        pages = {4091-4108},
          doi = {10.1093/mnras/stad2152},
archivePrefix = {arXiv},
       eprint = {2306.11791},
 primaryClass = {astro-ph.GA},
       adsurl = {https://ui.adsabs.harvard.edu/abs/2023MNRAS.524.4091B}
}

@ARTICLE{2008MNRAS.386..935F,
       author = {{Fraternali}, F. and {Binney}, J.~J.},
        title = "{Accretion of gas on to nearby spiral galaxies}",
      journal = {\mnras},
         year = 2008,
        month = may,
       volume = {386},
       number = {2},
        pages = {935-944},
          doi = {10.1111/j.1365-2966.2008.13071.x},
archivePrefix = {arXiv},
       eprint = {0802.0496},
 primaryClass = {astro-ph},
       adsurl = {https://ui.adsabs.harvard.edu/abs/2008MNRAS.386..935F}
}

@ARTICLE{2001MNRAS.328..726S,
       author = {{Springel}, Volker and {White}, Simon D.~M. and {Tormen}, Giuseppe and {Kauffmann}, Guinevere},
        title = "{Populating a cluster of galaxies - I. Results at z=0}",
      journal = {\mnras},
         year = 2001,
        month = dec,
       volume = {328},
       number = {3},
        pages = {726-750},
          doi = {10.1046/j.1365-8711.2001.04912.x},
archivePrefix = {arXiv},
       eprint = {astro-ph/0012055},
 primaryClass = {astro-ph},
       adsurl = {https://ui.adsabs.harvard.edu/abs/2001MNRAS.328..726S}
}

@ARTICLE{2025arXiv250800988D,
       author = {{Dubay}, Liam O. and {Johnson}, Jennifer A. and {Johnson}, James W. and {Roberts}, John D.},
        title = "{Challenges to the Two-Infall Scenario by Large Stellar Age Catalogs}",
      journal = {arXiv e-prints},
         year = 2025,
        month = aug,
          eid = {arXiv:2508.00988},
        pages = {arXiv:2508.00988},
          doi = {10.48550/arXiv.2508.00988},
archivePrefix = {arXiv},
       eprint = {2508.00988},
 primaryClass = {astro-ph.GA},
       adsurl = {https://ui.adsabs.harvard.edu/abs/2025arXiv250800988D}
}

@ARTICLE{2014ApJ...781L..31S,
       author = {{Snaith}, Owain N. and {Haywood}, Misha and {Di Matteo}, Paola and {Lehnert}, Matthew D. and {Combes}, Fran{\c{c}}oise and {Katz}, David and {G{\'o}mez}, Ana},
        title = "{The Dominant Epoch of Star Formation in the Milky Way Formed the Thick Disk}",
      journal = {\apjl},
         year = 2014,
        month = feb,
       volume = {781},
       number = {2},
          eid = {L31},
        pages = {L31},
          doi = {10.1088/2041-8205/781/2/L31},
archivePrefix = {arXiv},
       eprint = {1401.1835},
 primaryClass = {astro-ph.GA},
       adsurl = {https://ui.adsabs.harvard.edu/abs/2014ApJ...781L..31S}
}

@INPROCEEDINGS{2024IAUS..377..115N,
       author = {{Nidever}, David L. and {Gilbert}, Karoline and {Tollerud}, Erik and {Siders}, Charles and {Escala}, Ivanna and {Prieto}, Carlos Allende and {Smith}, Verne and {Cunha}, Katia and {Debattista}, Victor P. and {Ting}, Yuan-Sen and {Kirby}, Evan N.},
        title = "{The Prevalence of the {\ensuremath{\alpha}}-bimodality: First JWST {\ensuremath{\alpha}}-abundance Results in M31}",
    booktitle = {Early Disk-Galaxy Formation from JWST to the Milky Way},
         year = 2024,
       editor = {{Tabatabaei}, Fatemeh and {Barbuy}, Beatriz and {Ting}, Yuan-Sen},
       series = {IAU Symposium},
       volume = {377},
        month = jan,
        pages = {115-122},
          doi = {10.1017/S1743921323002016},
archivePrefix = {arXiv},
       eprint = {2306.04688},
 primaryClass = {astro-ph.GA},
       adsurl = {https://ui.adsabs.harvard.edu/abs/2024IAUS..377..115N}
}

@ARTICLE{2018MNRAS.474.3629G,
       author = {{Grand}, Robert J.~J. and {Bustamante}, Sebasti{\'a}n and {G{\'o}mez}, Facundo A. and {Kawata}, Daisuke and {Marinacci}, Federico and {Pakmor}, R{\"u}diger and {Rix}, Hans-Walter and {Simpson}, Christine M. and {Sparre}, Martin and {Springel}, Volker},
        title = "{Origin of chemically distinct discs in the Auriga cosmological simulations}",
      journal = {\mnras},
         year = 2018,
        month = mar,
       volume = {474},
       number = {3},
        pages = {3629-3639},
          doi = {10.1093/mnras/stx3025},
archivePrefix = {arXiv},
       eprint = {1708.07834},
 primaryClass = {astro-ph.GA},
       adsurl = {https://ui.adsabs.harvard.edu/abs/2018MNRAS.474.3629G}
}

@ARTICLE{2018MNRAS.477.5072M,
       author = {{Mackereth}, J. Ted and {Crain}, Robert A. and {Schiavon}, Ricardo P. and {Schaye}, Joop and {Theuns}, Tom and {Schaller}, Matthieu},
        title = "{The origin of diverse {\ensuremath{\alpha}}-element abundances in galaxy discs}",
      journal = {\mnras},
         year = 2018,
        month = jul,
       volume = {477},
       number = {4},
        pages = {5072-5089},
          doi = {10.1093/mnras/sty972},
archivePrefix = {arXiv},
       eprint = {1801.03593},
 primaryClass = {astro-ph.GA},
       adsurl = {https://ui.adsabs.harvard.edu/abs/2018MNRAS.477.5072M}
}

@ARTICLE{2021ApJ...921..106N,
       author = {{Nikakhtar}, Farnik and {Sanderson}, Robyn E. and {Wetzel}, Andrew and {Loebman}, Sarah and {Sharma}, Sanjib and {Beaton}, Rachael and {Mackereth}, J. Ted and {Poovelil}, Vijith Jacob and {Zasowski}, Gail and {Bonaca}, Ana and {Martell}, Sarah and {J{\"o}nsson}, Henrik and {Faucher-Gigu{\`e}re}, Claude-Andr{\'e}},
        title = "{New Families in our Solar Neighborhood: Applying Gaussian Mixture Models for Objective Classification of Structures in the Milky Way and in Simulations}",
      journal = {\apj},
         year = 2021,
        month = nov,
       volume = {921},
       number = {2},
          eid = {106},
        pages = {106},
          doi = {10.3847/1538-4357/ac1a10},
archivePrefix = {arXiv},
       eprint = {2104.08394},
 primaryClass = {astro-ph.GA},
       adsurl = {https://ui.adsabs.harvard.edu/abs/2021ApJ...921..106N}
}

\begin{appendix}
\section{GMM disc components}
\label{app:GMM_discs}
\vspace{0.5cm}

\begin{figure}[h]
    \centering
    \includegraphics[width=0.95\columnwidth]{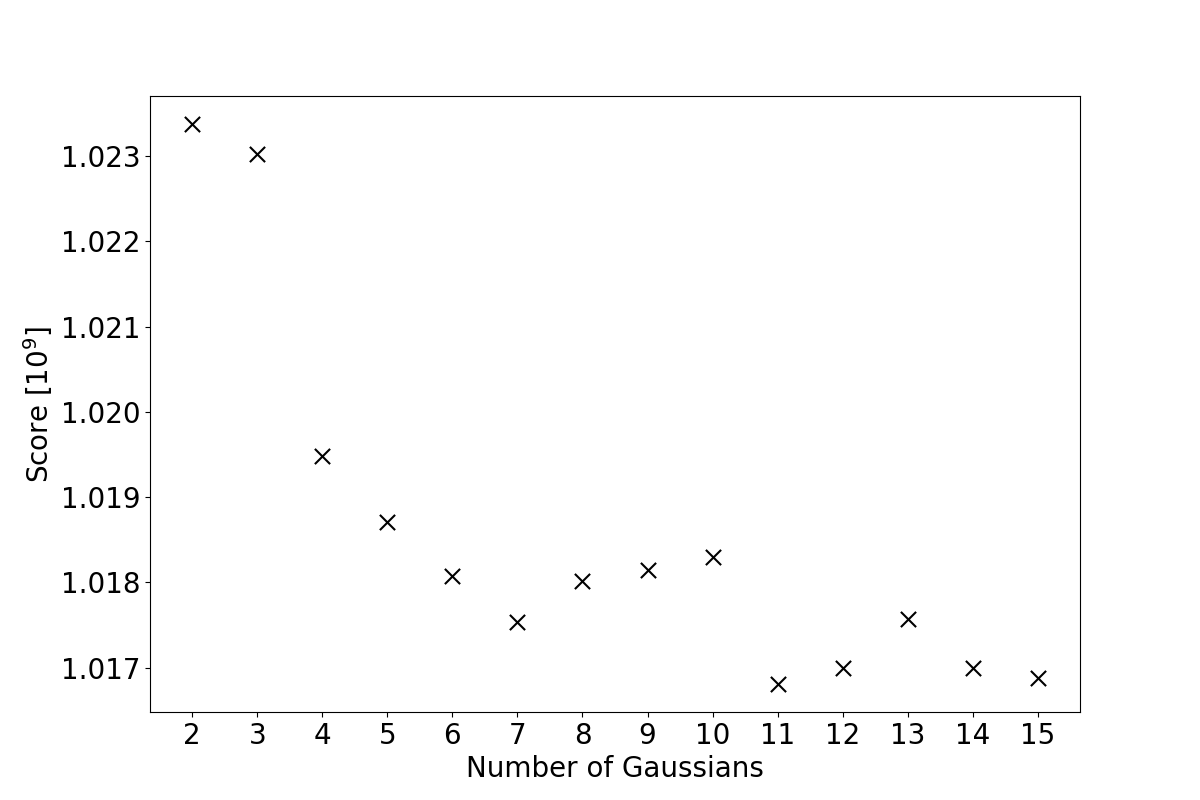}
    \caption{Bayesian Information Criterion as a function of the number of Gaussian components in the GMM analysis of 3D cylindrical velocities and stellar ages within $R < 40$ kpc and $|z| < 10$ kpc in the {\sc Romeo} simulated galaxy.}\label{fig:BIC}
\end{figure}

In Fig.~\ref{fig:BIC}, we present the Bayesian Information Criterion (BIC) values as a function of the number of Gaussian components used to model the joint probability density of 3D cylindrical velocities and ages of star particles within $R<40$ kpc and $|z|<10$ kpc in {\sc Romeo}. The decrease in BIC values slows noticeably beyond four components. Moreover, increasing the number of components further introduces nonphysical clusters, such as artificial separations in radial velocity into distinct positive and negative groups.

\begin{figure*}
    \centering
    \includegraphics[width=0.95\textwidth]{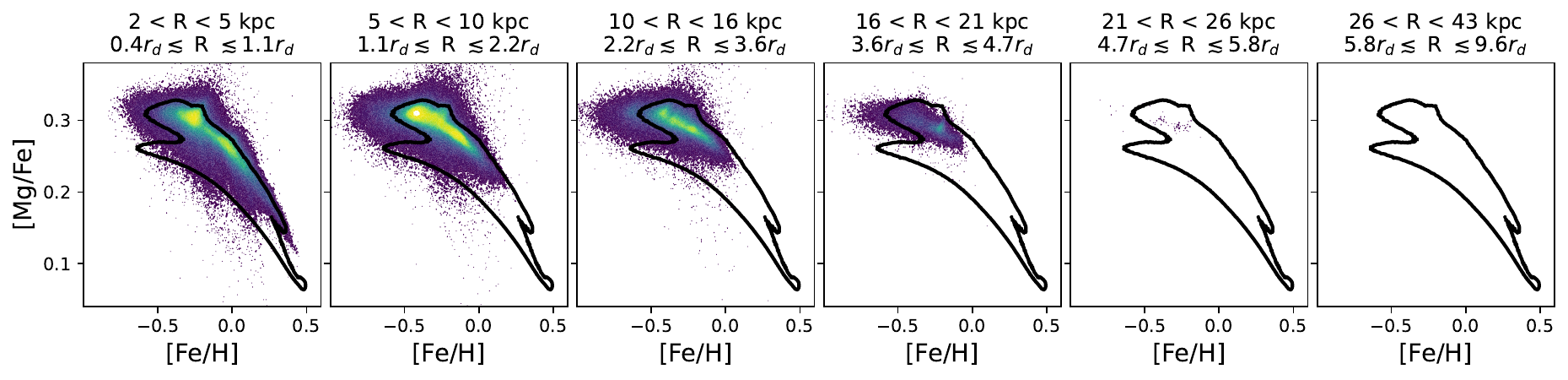}
    \includegraphics[width=0.95\textwidth]{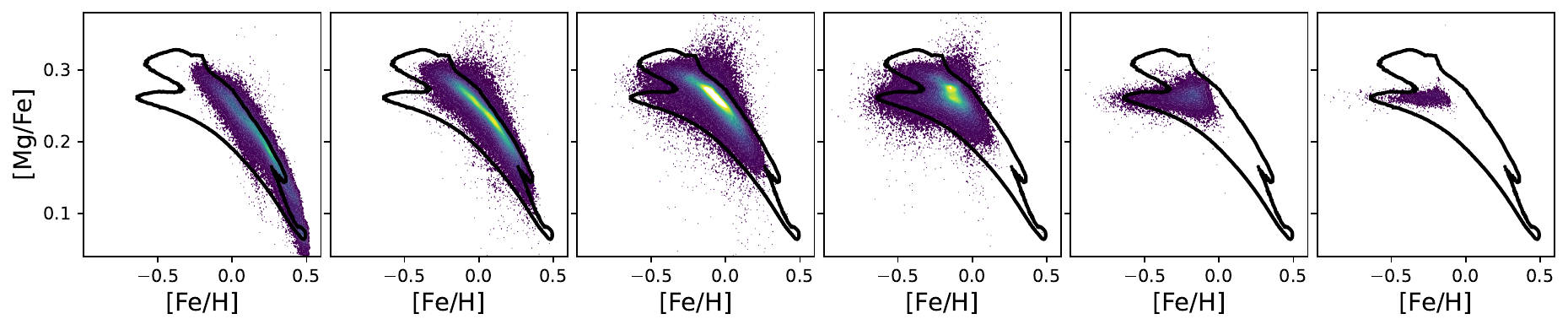}
    \includegraphics[width=0.95\textwidth]{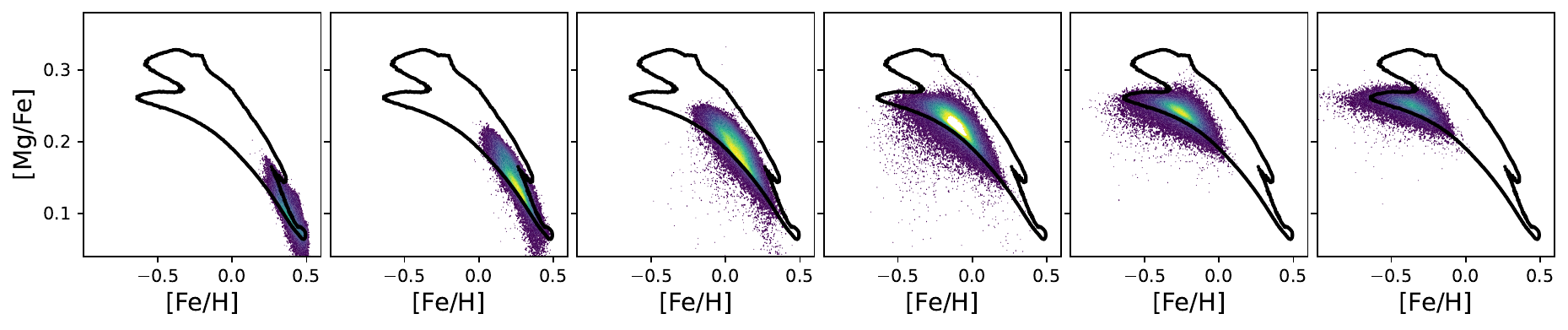}
      \captionof{figure}{Number density for the high-$\alpha$ (top), bridge (middle) and low-$\alpha$ (bottom) discs. The 2D histograms in each row are scaled by the largest amplitude among the six subplots in that row. In this way, the galactocentric distances at which each disc component is most prominent can be seen.}\label{fig:global_disc_radial_components}
\end{figure*}

Fig.~\ref{fig:global_disc_radial_components} shows the two-dimensional number density of high-$\alpha$, bridge and low-$\alpha$ stars at different annular rings with $|z| < 1$ kpc

\section{Gaia-APOGEE data preparation and selection criteria}
\label{app:GaiaApogee}
\vspace{0.5cm}

Gaia DR3-provided astrometry and radial velocities, coupled with stellar distance estimates from GSP-Phot~\citep{gspphot}, were used to transform the sample to the Galactocentric frame and cylindrical coordinates. This was done by adopting the Sun’s orbital parameters from~\cite{bay_gaia} and using the transformation procedure and the Gaia-tools repository\footnote{\href{https://github.com/HEP-KBFI/gaia-tools}{https://github.com/HEP-KBFI/gaia-tools}} described therein. Finally, counter-rotating stars were removed and the sample was radially restricted to a Galactocentric range of $R \in [4, 16] \rm \, kpc$ and vertically within $\pm 2 \rm \, kpc$.

As for the APOGEE DR17 data, following~\cite{fernandez-alvar_metal-poor_2024}, we selected only stars with ASPCAPFLAG bits 14–42 equal to 0, STARFLAG bits 1–26 equal to 0, and EXTRATARG bits 2 and 4 equal to 0. We also excluded those objects with the {\it apogee1\_target1}, {\it apogee1\_target2}, {\it apogee2\_target1}, and {\it apogee2\_target2} classifying them as clusters, streams, dwarf galaxies, sky, telluric, binaries, radial velocity variables, the bar, extended objects, and the TriAnd, GASS, and A13 disc structures. In addition, we select only stars with 3500 < T$_{\rm eff}$ [K] < 6500 and 1 < $\log g$ [dex] < 3.5.

\section{Radial migration}
\label{app:radial_migration}
\vspace{0.5cm}

In Fig.~\ref{fig:radial_migration}, we show the difference between cylindrical radius of low-$\alpha$ disc stars at formation and at the snapshot prior to bar formation. The majority of the stars lie close to the zero horizontal line.
We observe that, although there is an apparent average inward motion for galactocentric radii greater than 25 kpc, the median line overestimates this effect for two reasons. First, there is a statistical bias because the conditioning on the formation radius skews the median toward lower final radii as a consequence of the intrinsic dispersion of the formation radius. Second, even in the absence of radial migration, we expect some stars to move slightly inward and others slightly outward. However, in the outer regions, since the galaxy (or disc) is not infinite, there are no stars that can arrive from beyond the edge to balance the distribution. This lack of symmetry results in a biased distribution, further reinforcing the apparent inward trend.

\noindent\begin{minipage}{\columnwidth}
    \centering
    \includegraphics[width=0.7\columnwidth]{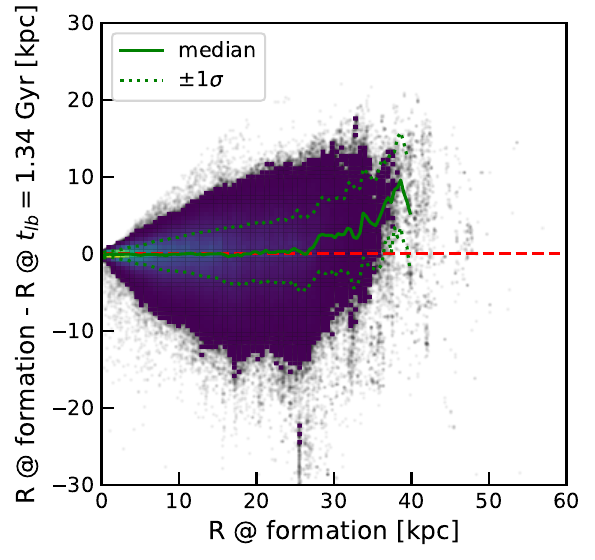}
      \captionof{figure}{Difference between the cylindical radius at formation and at the snapshot prior to bar formation versus formation radii for low-$\alpha$ disc stars. The green solid line shows the median trend with the corresponding 1$\sigma$ scatter shown by the dashed green lines.
      }
    \label{fig:radial_migration}
\end{minipage}

\section{Calculation of gas flow rates}
\label{app:gas_flow_rates}
\vspace{0.5cm}

Following~\citet{2019MNRAS.489.4233M}, for each snapshot, the flux through a surface was computed as
\begin{equation}
    \dot{\rm M} = \frac{\sum v_n m}{\Delta}, 
\end{equation}
where $m$ and $v_n$ are the gas cell mass and its velocity component perpendicular to the surface, and the sum runs over gas cells within a slice of thickness $\Delta=0.3\,\rm kpc$\footnote{Varying $\Delta$ within reasonable limits does not affect the results.}. In particular, for each annular ring defined by $R_{\min}<R<R_{\max}$ and $|z|<2\,\mathrm{kpc}$, vertical fluxes were measured through the top and bottom surfaces by selecting cells within $R_{\min}<R<R_{\max}$ and $z\in[\pm 2-\Delta/2, \pm 2+\Delta/2]$. Cells with $v_z\,z>0$ ($v_z\,z < 0$) were counted as outflows (inflows).
Additionally, radial fluxes were measured through the outer cylindrical surface by selecting gas cells with $R\in[R_{\max}-\Delta/2, R_{\max}+\Delta/2]$ and $|z|<2\,\mathrm{kpc}$, with cells with $v_R\,R >0$ ($v_R\,R<0$) classified as outflows (inflows). 

At each snapshot, the concentric annular cylinders are aligned with the coordinate system defined by the galaxy's principal axes, which are calculated as the eigenvectors of the moment-of-inertia tensor. This tensor was determined using the youngest 25\% of star particles that, within 10 kpc, comprise 90\% of the total stellar mass~\citep{wetzel_public_2023}. While an alternative approach could be to align the coordinate system with the total angular momentum of the stellar disc, as done in \cite{2019MNRAS.482.3089N}, the current choice of alignment ensures that primarily disc particles are selected. Furthermore, for a disc-like configuration, the principal axis of the moment-of-inertia tensor generally aligns with the total angular momentum vector.

\end{appendix}
\end{document}